\pdfoutput=1

\documentclass[twocolumn,showpacs,preprintnumbers,amsmath,amssymb]{revtex4}
\usepackage{graphicx}% Include figure files
\usepackage{dcolumn}% Align table columns on decimal point
\usepackage{bm}% bold math

\begin{document}

\preprint{APS/123-QED}

\title{Matter-wave cavity gravimeter}% Force line breaks with \\

\author{F. Impens$^{1}$, P. Bouyer$^{2}$ and Ch. J. Bord\'{e}$^{1,3}$}
\affiliation{$^{1}$ SYRTE, CNRS UMR 8630, Observatoire de Paris, 61 avenue de l'Observatoire 75014 Paris\\
$^{2}$ Laboratoire Charles Fabry de l'Institut d'Optique, CNRS UMR 8501, 91403 Orsay Cedex, France\\
$^{3}$ Laboratoire de Physique des Lasers, Institut Galil\'{e}e,
CNRS UMR 7538, Universit\'{e} Paris Nord, F-93430
Villetaneuse,France\\}

\date{\today}% It is always \today, today,

\begin{abstract}
We propose a gravimeter based on a matter-wave resonant cavity
loaded with a Bose-Einstein condensate and closed with a sequence
of periodic Raman pulses.  The gravimeter sensitivity increases
quickly with the number of cycles experienced by the condensate
inside the cavity. The matter wave is refocused thanks to a
spherical wave-front of the Raman pulses. This provides a
transverse confinement of the condensate which is discussed in
terms of a stability analysis. We develop the analogy of this
device with a resonator in momentum space for matter waves.
\end{abstract}


\maketitle

%\section{Introduction}

The realization of gravimeters loaded with cold atomic clouds has
drastically increased the accuracy of the measurement of
gravitational acceleration to a few parts per
billion~\cite{Chu99,Chu01}. The recent obtention of
quasi-continuous atom lasers now opens new perspectives for
gravito-inertial sensors with the possibility to load these
devices with a fully coherent and collimated matter source instead
of the incoherent cold atomic samples used so far. In this paper
we investigate a gravimeter based on a resonant matter-wave cavity
loaded with a Bose-Einstein condensate.  The condensate is
stabilized in momentum space thanks to a sequence of periodic
``mirror pulses''  consisting in velocity-sensitive double Raman
$\pi$-pulses. Between the pulses, the optical potential is shut
off and the condensate experiences a pure free fall. Other
measurements of the acceleration of gravity have been proposed,
based on the Bloch oscillations of an atomic cloud in a standing
light wave~\cite{Biraben05,Borde87} or on the bouncing of a cloud
on an evanescent-wave optical cavity~\cite{WalDalCoh91,Bouyer93}.
In our setup, we can minimize parasitic diffusions
processes~\cite{Landragin96} which may kick atoms out of the
cavity and thus limit the lifetime $T_{cav}$. The expansion of the
atomic cloud, whose size quickly exceeds the diameter of the
mirror, usually limits the number of bounces in the
cavity~\cite{Bouyer93}. As in~\cite{WalDalCoh91,Bouyer93}, we
circumvent this problem by using a ``curved mirror'' which
refocuses periodically the
condensate. %so that it keeps finite transverse dimensions. The
%cloud expansion thus does not limit fundamentally the lifetime
%$T_{cav}$.
This gives a very promising sensitivity for the proposed
gravimeter, which increases as $T_{cav}^{3/2}$ with the atom
interrogation time $T_{cav}$ as in standard atom
gravimeters~\cite{Chu99}.

\section{Determination of the acceleration of gravity: principle of the measurement}
\label{sec:principle_gravity_determination}

Following up our approach in~\cite{Impens06}, we present in this
section a first simple description of the proposed matter-wave
cavity and give a heuristic analysis of its performance as a
gravimeter.

%The period of the sequence of "mirror pulses" is the key-parameter
%of our experiment. A fine tuning of this period as to minimize the
%losses of the cavity provides a determination of

\subsection{Principle of the experiment}

The principle is to levitate a free falling atomic sample by
providing a controllable acceleration mediated by a coherent
atom-light interaction. Radiation pressure could provide
levitation, but the resulting force is not precisely tunable if
tied to incoherent spontaneous emission processes. A better choice
to provide this acceleration is thus a series of vertical Raman
pulses. Indeed these pulses impart coherently a very well defined
momentum to a collection of atoms~\cite{Chu94}. A sequence of
Raman pulses of identical effective wave vector $\mathbf{k}$
interspaced with a duration $T$ gives an acceleration to an atomic
cloud which is monitored by the choice of $T$. Levitation occurs
when the sequence of vertical Raman pulses compensates, on
average, the action of gravity. This stabilization is obtained
thanks to a fine-tuning of the period between two pulses: after a
fixed time, one observes a resonance in the number of atoms kept
in the cavity for the adequate period $T_0$. The atomic cloud is
then well stabilized, and the average Raman acceleration equals
the gravitational acceleration. Knowing the period $T_0$, one can
infer the corresponding Raman acceleration and thus the gravity
acceleration $g$. The ratio $\hbar k /m$ can be simultaneously
determined using the resonance condition
of the Raman mirrors.\\

\subsection{Description of the cavity}

As displayed in Fig.1, the atomic sample, initially at rest in the
lower state $a$, is dropped. After a free fall of duration $T/2$,
during which the sample acquires a momentum $g T / 2$, we shine a
first Raman $\pi$ pulse of $2$ counterpropagating lasers with
respective wave vectors $\mathbf{k_{down}}=\mathbf{k_1}$ and
$\mathbf{k_{up}}=\mathbf{k_2}$ and respective frequencies
$\omega_{down}=\omega_1$ and $\omega_{up}=\omega_2$. This brings
the atom from the internal state $a$ to an internal state $b$ with
a momentum transfer $2 \hbar k = k_2 -k_1$. Then we shine a second
Raman $\pi$ pulse with $\mathbf{k_{down}}=\mathbf{k_4}$ and
$\mathbf{k_{up}}=\mathbf{k_3}$, with respective frequencies
$\omega_{down}=\omega_4$ and $\omega_{up}=\omega_3$ $( \omega_3 -
\omega_4 \sim \omega_2 - \omega_1)$. This pulse brings the atomic
internal state back to state a with an additional momentum
transfer $2 \hbar k$. If the two successive pulses are
sufficiently close, this sequence acts as a single coherent
``mirror pulse'' which keeps the same internal state a and
modifies the atomic momentum by $4 \hbar k$. In particular, if the
initial momentum is $- 2 \hbar k$, the mirror simply inverts the
velocity. This ``mirror pulse'' is velocity-sensitive: it reflects
only the atoms whose vertical momentum belong to a tiny interval
around a specific value $p_0$. Thus, in order to bounce several
times, the atoms should have the same momentum $p_0$ immediately
before each ``mirror pulse''. This implies a resonance
condition~(\ref{eq:resonanceT}) on the period between two pulses.
The adequate momentum $p_0$ is set by the energy conservation
during the pulse and fulfills the resonance
condition~(\ref{eq:resonanceBragg}).
%The average altitude and
%momentum of the condensate as a function of time are represented
%on Fig. \ref{fig:altitude_momentum}.
As depicted in Fig.~\ref{fig:altitude_momentum}, if the resonance
condition~(\ref{eq:resonanceT}) is satisfied, the sample will have
a periodic trajectory in the momentum space. It is this
periodicity of the atomic momentum for an adequate time-spacing
$T_0$ of two successive ``mirror pulses'' which yields the picture
of a matter-wave cavity in momentum space.
Fig.~\ref{fig:phase_space_diagram} represents an energy-momentum diagram of the atomic sample during a cavity cycle.\\

\begin{figure}[htbp]
\begin{center}
\includegraphics[width=5.2cm]{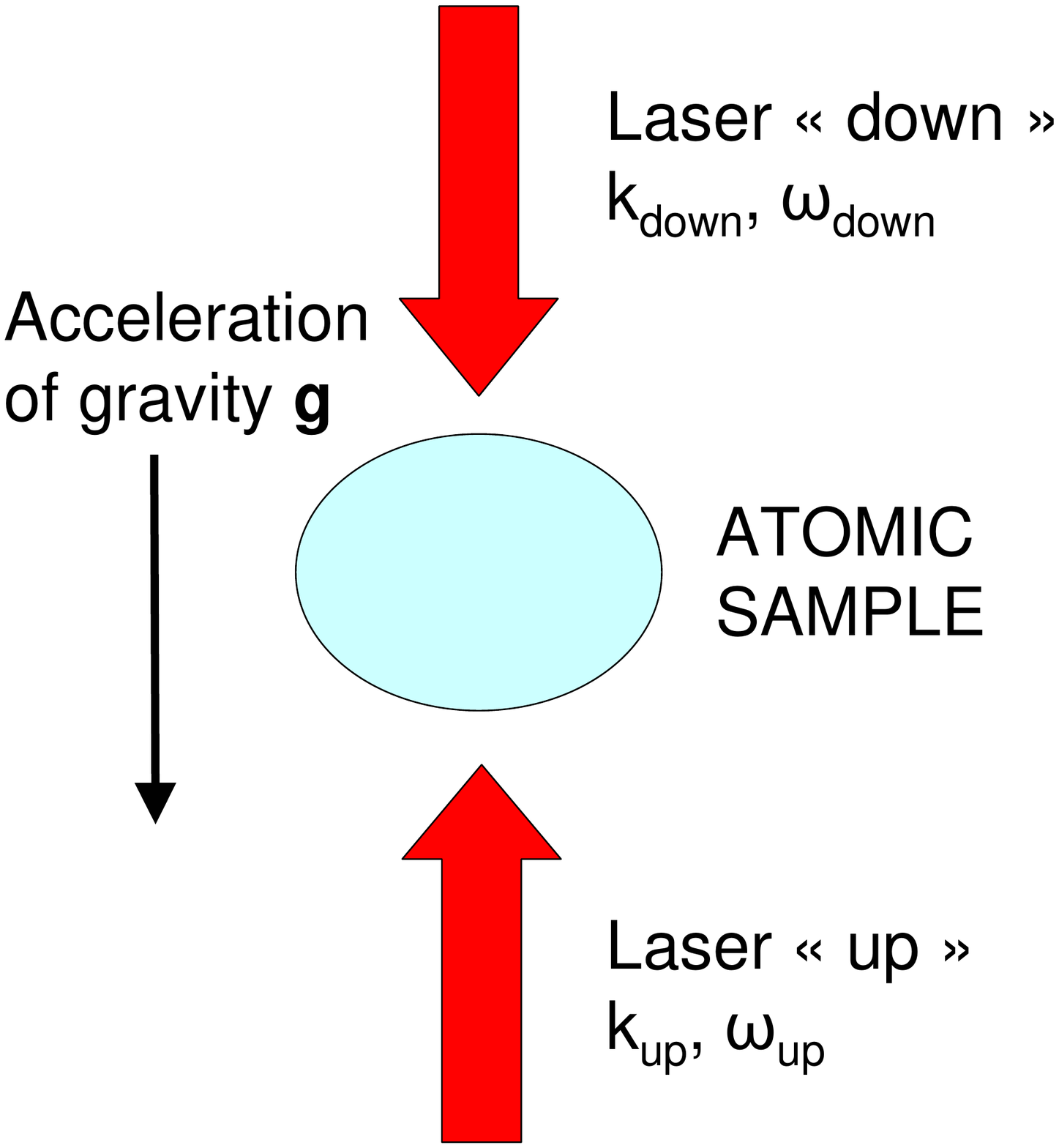}
\end{center}
\caption{Setup description.} \label{fig:setup}
\end{figure}

\begin{figure}[htbp]
\begin{center}
\includegraphics[width=7.5cm]{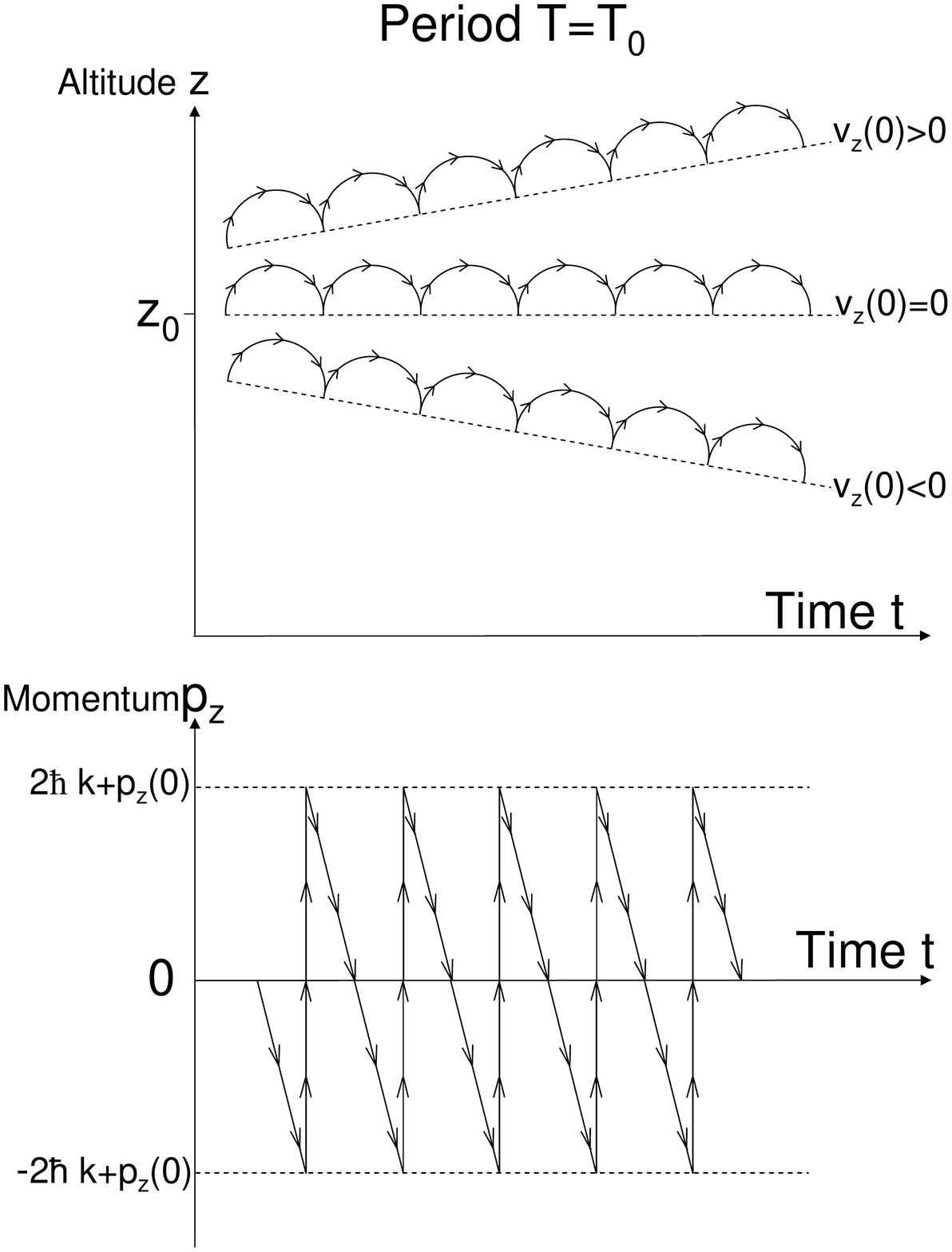}
\end{center}
\caption{Altitude and momentum of the atom sample as a function of
time for the resonant period $T=T_0$ and for different initial
velocities. The atom cloud always undergoes a periodic motion in
momentum space even with a nonzero initial velocity. The figure
illustrates three different cases of stable cavity in momentum
space.} \label{fig:altitude_momentum}
\end{figure}

\begin{figure}[htbp]
\begin{center}
\includegraphics[width=7.5cm]{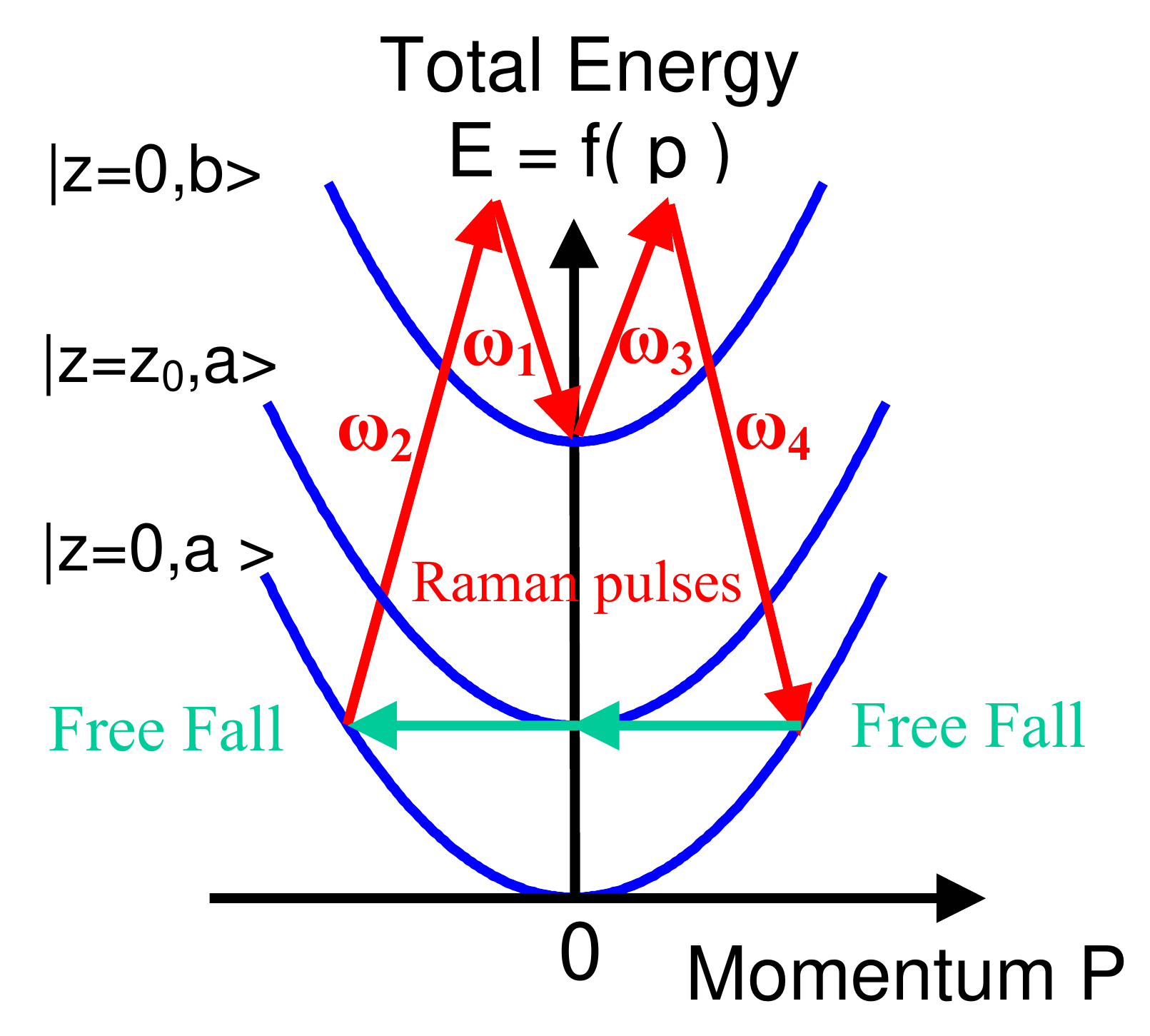}
\end{center}
\caption{Energy-momentum diagram of the condensate. $z_0$
represents the height from which atoms are dropped.}
\label{fig:phase_space_diagram}
\end{figure}

\subsection{Resonance Conditions}

For the resonant period $T_0$, the momentum kick imparted during
each ``mirror pulse'' is equal to the momentum acquired through
the free fall between two such pulses:
\begin{equation}
T_0 = \frac {4 \hbar k}  {m g} \label{eq:resonanceT}
\end{equation}
%In a sequence of "mirror pulses" with period $T_0$, the rebound of the
%condensate on the "mirror pulse" is elastic and it brings back the
%condensate in its initial state, at rest at the altitude $z_0$.
%The condensate may then start again an other
%free-fall - reflection cycle.
%When a sequence of "mirror pulses" is applied with a period $T_0$,
%the condensate describes a closed trajectory in the momentum space
%and goes back periodically in its initial state, as is depicted on
%Fig.~\ref{fig:altitude_momentum}
If the period $T$ differs from $T_0$, the atomic cloud takes an
acceleration $a= 4 \hbar k / m \left( 1 / T -  1 / T_0 \right)$.
The average speed resulting from this acceleration drifts the
momentum of the atoms from the optimum value $\mathbf{p_0}=-2\hbar
k \mathbf{u_z}$ satisfying the Bragg
condition~(\ref{eq:resonanceBragg}) associated with elastic energy
conservation. This drives the atoms progressively out of resonance
and a part of the cloud will not be reflected by the ``mirror
pulses''. When $T$ differs from $T_0$, we thus observe a drift in
position and momentum as well as a leakage of the condensate. It
should be noted that if $T=T_0$, a non-zero initial velocity does
not reduce fundamentally the number of bounces of an atomic sample
(Fig.~\ref{fig:altitude_momentum}). Indeed, only the periodicity
of the momentum space trajectory matters, and provided that one
adjusts the Raman detuning to account for the shift in the
momentum $p_0$, the sample can still be reflected several times.
The atomic sample merely drifts with a constant average vertical
velocity, the only limitation being then the finite size of the
experiment. Thanks to this flexibility in the initial velocity,
the resonance observed is robust to an imperfect timing of the
trap shutdown.\\

Bragg resonance conditions express the energy conservation during
the pulses: %If the atom is at rest between the two Raman pulses, the adequate detunings are opposite:
\begin{eqnarray}
\label{eq:resonanceBragg} \omega_1 + \omega_{b}+\delta_{AC}+ \frac
{(\textbf{p}+2\hbar \textbf{k})^2} {2m \hbar} =
\omega_2+\omega_{a}+\frac {\textbf{p}^2}
{2 m \hbar} \nonumber \\
 \omega_4 + \omega_a + \frac {(\textbf{p}+4 \hbar \mathbf{k})^2}
{2 m \hbar} = \omega_3 + \omega_{b}+\delta_{AC}+ \frac
{(\textbf{p}+2 \hbar \mathbf{k})^2} {2 m \hbar}
\end{eqnarray}
%The sign "+" and "-" on the right hand side stand for the Bragg
%conditions on respectively the first and second Raman pulses.
$\textbf{p}$ is the matter-wave average momentum immediately
before the first Raman pulse, and $\delta_{AC}$ is the light
shift. Thanks to this second set of conditions, which directly
impacts the reflection coefficient, the ``mirror pulses'' act
directly as the probe of the resonant time-spacing $T_0$ expressed
in~(\ref{eq:resonanceT}). When the atomic sample is dropped
without initial speed, for a nearly resonant period $T \simeq
T_0$, the first pulse brings the sample at rest, so that both
pulses play a symmetric role. If one does the replacements
$\omega_1 \rightarrow \omega_3$ and $\omega_2 \rightarrow
\omega_4$, the mismatch in the two Bragg conditions is then equal
in absolute value, yielding identical reflection coefficients for
both pulses. One can then consider that the two Bragg conditions
merge into a single one. %This will alleviate the notations.
We assume from now on that the atomic cloud has no initial
velocity \footnotemark[1] \footnotetext[1]{Bose Einstein
condensates can be brought at rest very accurately and are thus
well suited for our system.},
but the extension to the general case is straightforward.\\
%One could therefore
%choose to scope either the period $T$ or the Raman detuning
%$\omega_2-\omega_1$, the other parameter being only approximately
%its optimum value, to observe a resonance in the number of atoms
%present in the cavity after a certain number of cycles and
%determine the gravity $g$.
The two conditions (\ref{eq:resonanceT}) and
(\ref{eq:resonanceBragg}) must be satisfied
 to ensure the resonance of the matter-wave
 cavity. Nonetheless, condition (\ref{eq:resonanceT}) on the period is
much more critical than the Bragg condition
(\ref{eq:resonanceBragg}). Indeed, a slight shift in the period
$T$ from its optimum value $T_0$ implies for the condensate an
upward or downward acceleration. The increasing speed acquired by
the atoms will generate, through Doppler shifting, a greater
violation of the Bragg resonance condition and thus greater losses
at each ``mirror pulse''. Conversely, a mismatch in the detuning
$\omega_2-\omega_1$ with the adjustment $T=T_0$ will only induce
constant losses at each cycle. The observation of a resonance in
the number of atoms, when one scans the period between two Raman
pulses, is thus very sharp and robust to an imperfect adjustment
of the Raman detuning. Consistently, we choose to determine the
acceleration of gravity $g$ through condition
(\ref{eq:resonanceT}). Fig. \ref{fig:resonance_after 1 10 50
cycles} sketches the number of atoms in the cavity as a function
of the period $T$ after different numbers of cycles with a
detuning matching perfectly condition (\ref{eq:resonanceBragg}).
We observe that the resonance in $T$ becomes sharper
as the number of cycles increases.\\

\begin{figure}[htbp]
\begin{center}
\includegraphics[width=7.5cm]{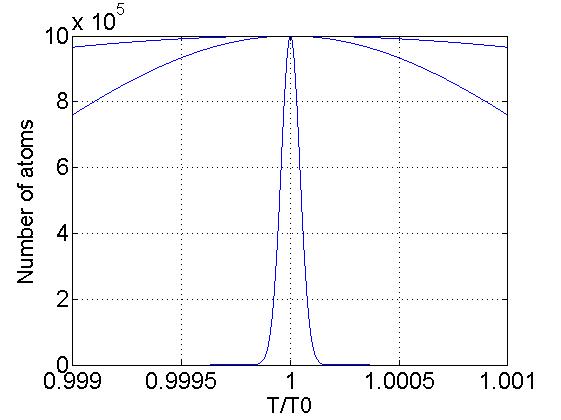}
\end{center}
\caption{Number of atoms in the cavity after 1,10,50 cycles as a
function of the ratio $T/T_0$ $(T_0 \simeq 3.8 \: \mbox{ms})$. We
took $\Omega_0= 2 \pi \times 5 \times 10^3 \: \mbox{Hz}$. The
number of
 atoms initially present in the
cavity was fixed to be $N=10^6$.}\label{fig:resonance_after 1 10
50 cycles}
\end{figure}

\subsection{Expected Sensitivity}

Let us derive the resonance figure associated with the period $T$.
We characterize the mismatch in the Bragg condition by an
``off-Braggness'' parameter $y(\mathbf{p})$ function of the
average momentum of the atoms~\cite{BordeBerman}:
\begin{equation}
\label{eq:OffBraggparam1} y(\mathbf{p}) = \frac {1} {2 \Omega_0}
[\hbar\mathbf{k}^{2} /  2 m+2 \mathbf{p}
 \cdot \mathbf{k} / m - (\omega_1- \omega_2 - \omega_{ba}-\delta_{AC})]
\end{equation}
where $\Omega_0$ the effective Rabi frequency of the Raman pulse
and $\mathbf{p}$ the momentum immediately before the first pulse..
The Raman detuning is adjusted to be resonant if $\mathbf{p}=p_0
\mathbf{u_z}=-2 \hbar k \mathbf{u_z}$, so we adjust the detuning
to match $y(-2 \hbar k \mathbf{u_z})=0$:
\begin{equation}
\label{eq:OffBraggparam2} y(\mathbf{p}) =  \frac {(p_z+ 2\hbar k)
k}
 {2 m  \Omega_0} = \frac {(p_z+m g T_0 /2) k}
 {2 m  \Omega_0}
\end{equation}
In the remainder of this section, we focus on the vertical
component of momentum which we note $p$ to alleviate the
notations. Because of the mismatch in the Bragg condition, only a
fraction $\rho(p)$ of the atomic cloud will then be transferred
during the first Raman pulse:
\begin{equation}
\rho(p)= \frac {\sin^2 \left(\frac {\pi} 2 \sqrt{1+ y(p)^2}
\right)} {1+ y(p)^2}
\end{equation}
For the second Raman pulse, the mismatch is equal in value and
opposite in sign, so the same fraction of atoms undergoes the
second transition. The reflection coefficient of the ``mirror
pulse'' is then simply the product of these values:
\begin{equation}
\label{eq:reflection coefficient} R(p)= \frac {\sin^4 \left( \frac
{\pi} 2 \sqrt{1+ y(p)^2} \right)} {(1+ y(p)^2)^2}
\end{equation}
The part of the cloud which is not reflected will simply go on a
free fall and have an off-Braggness parameter of $y(p-m g T)$ for
the next ``mirror pulse''. We will adopt experimental parameters
such that $y(m g T_0) \gg 1$, so that non-reflected atoms are
insensitive to subsequent Raman transitions and can be considered
as expelled from the cavity.

\begin{figure}[htbp]
\begin{center}
\includegraphics[width=7.5cm]{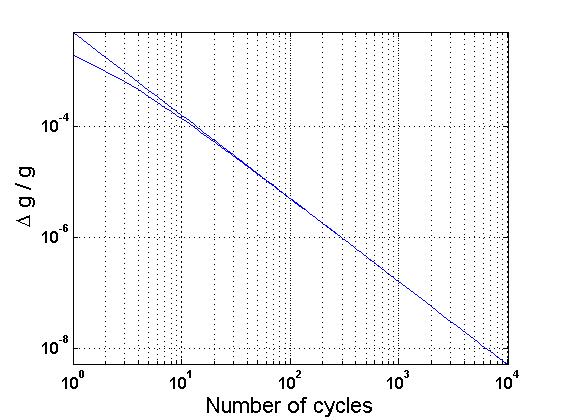}
\end{center}
\caption{Error on $\Delta g / g$ as a function of the number of
cycles, with Raman pulses of Rabi pulsation $\Omega_0= 2 \pi
\times 5 \times 10^{3} Hz$ and a detection threshold of
$\epsilon=10^{-3}$. The lower curve is the actual sensitivity
based on a simulation with a condensate of infinitely narrow
momentum distribution. The upper curve is the analytic formula
(\ref{eq:relative_error}).}\label{fig:width_function cycle number}
\end{figure}

The average momentum acquired by the atomic cloud results from a
competition between the gravitational acceleration and the kicks
of the ``mirror pulses''. Given a period $T$ for the sequence, the
average vertical momentum of the cloud immediately before the n-th
``mirror pulse'' is simply:
\begin{eqnarray}
\label{eq:momentum_at_pulse_n} p_n & = & - m g T/2+ (n-1) \times m
g (T_0 - T )
\end{eqnarray}
The remaining fraction of the cloud after $n$ cycles is thus:
\begin{equation}
R(T) = R(p_1)...R(p_n)
\end{equation}
We expand this expression in
Appendix~\ref{sec:appendix_sensitivity computation} for nearly
resonant pulses. We have assumed that a relative variation
$\epsilon$ of the condensate population can be tracked
experimentally. This computation shows that the error in the
determination of the gravity acceleration can be less than:
\begin{equation}
\label{eq:relative_error} \frac {|\Delta {g}|} {g} \leq \sqrt
{\frac {3} {8}} \frac {1} {\hbar k^2} \left( \frac {\Omega_0
\sqrt{-\log(1-\epsilon)}} {n^{3/2}} \right)+ |\frac {\Delta v_r}
{v_r}|
\end{equation}
where $v_r= \hbar k /m$ is a recoil velocity which can be measured
independently or, as stated before, directly from resonance
condition~(\ref{eq:resonanceBragg})~\cite{LeCoq01}. This velocity
has been determined with an accuracy as good as a few $10^{-9}$
for Cs~\cite{Chu93,Wicht02} and Rb atoms~\cite{Biraben04},
ultimately limits the performance of our gravimeter. Fig.
\ref{fig:width_function cycle number} displays the relative error
in the determination of the acceleration of gravity obtained from
a numerical simulation with the reflection coefficient
(\ref{eq:reflection coefficient}). It shows a very good agreement
with the analytic expression (\ref{eq:relative_error}) after about
$10$ cycles.

 Formula (\ref{eq:relative_error}) thus yields a sensitivity which scales
as $n^{3/2}= ( T_{cav} / T_0 )^{3/2}$, where $T_{cav}$ is the
interrogation time of the atoms and $T_0$ the duration of a cycle
at resonance. We thus obtain the expected improvement of the
sensitivity with the atom interrogation time. This is normal since
the accuracy of the measurement also increases with the
selectivity of the Raman pulses. The mirror pulses should thus be
as long as possible to act as efficient atom velocity probes.
Ideally, each pulse should last half of the optimal period $T_0$.
The atom sample would then fall in a continuous light field. This
setup ressembles that of Clad\'{e} et al.~\cite{Biraben05}, except
that here the atomic sample is interacting with a travelling wave,
and that in addition to condition~(\ref{eq:resonanceT}) we have a
double
Raman condition~(\ref{eq:resonanceBragg}).\\
%In Section
%\ref{sec:interactions effects} we adapt the ABCD matrix formalism
%of matter-wave propagation to take into account interactions
%effects.

In the previous description, one could in principle maintain the
sample for an arbitrary long time inside the cavity provided that
the two resonance conditions are fulfilled, yielding an extremely
accurate measurement of the gravitational acceleration. In fact,
even at resonance, systematic losses occur that limit the number
of condensate cycles. Practically, one can hope to keep a
significant number of atoms in the cloud up to a certain number of
$n$ cycles, which reflects the sample lifetime $T_{cav}=n T_0$ at
resonance. These losses at resonance have several
origins.\\

First, they can result from imperfect recoil transfers due to
residual fluctuations in the Raman lasers intensities. Raman
impulsions can be very effective though, since these transfers
have been realized with an efficiency as high as
$99.5\%$.%~\cite{Biraben04}.

Second, parasitic diffusions may eject atoms from the cavity.
These processes, such as absorption followed by spontaneous
emission, can be made arbitrarily small by using far-detuned Raman
pulses. This limitation is thus essentially of technical nature.

Third, the cloud expansion can drive the atomic sample out of the
Raman lasers. Indeed, if the cloud is not refocused, its
transverse size exceeds quickly the diameter of the Raman lasers.
 The atoms would then be limited to a few cycles in the cavity. It is thus essential to involve a
mechanism that stabilizes transversally the atomic cloud. We
expose in Section \ref{sec:refocusing} two ways to focus the
atomic sample.

%The third loss factor is the imperfect stabilization of the cloud
%expansion. If the focusing is not effective enough, a part of the
%cloud may leak out of the trap. We believe that focusing could be
%more effective with a condensate.

\section{ABCD analysis of the matter-wave cavity}
\label{sec:condensate_propagation}
%We consider loading the gravimeter with a cigar-shaped condensate
%whose long axis $\mathbf{x}$ lies on the horizontal plane.

In the following, we will calculate explicitly the evolution of an
atom sample in our gravitational cavity using the ABCD matrix
formalism~\cite{BordeHouches,BordeTheortool2001,BordeMetrologia2002}.
In this section, we assume that the atom density after the initial
free fall is sufficiently low to make the effect of interactions
negligible during the subsequent bounces.

\subsection{Description of the atomic sample}
\label{subsec:condensate description}

We will restrict ourselves to a Bose-Einstein condensate which, in
the Hartree-Fock approximation, can be described by a macroscopic
wavefunction $\Psi(\mathbf{r},t)$. After the initial free-fall,
the evolution follows the linear partial differential equation:
\begin{equation}
i \hbar \partial_t | \Psi(\mathbf{r},t) \rangle =
[p^2/2m+H_{\mbox{G}}] | \Psi(\mathbf{r},t) \rangle
\end{equation}
where $H_{\mbox{G}}$ is the Hamiltonian associated with
gravito-inertial effects. From the linearity of this equation, we
can deduce the time evolution of any arbitrary wave function from
the propagation of a complete set of functions. As explained in
Appendix~\ref{sec:appendix_generating function method},
  the propagation of such a basis, the Hermite-Gauss modes $H_{lmn} (\mathbf{r})$, can be extracted from the
propagation of a generating Gaussian wave
function~\cite{BordeMetrologia2002}:
\begin{eqnarray}
\label{eq:generatinggaussian} & \Psi_{\overrightarrow{\alpha}}
(\mathbf{r}) =  \frac {1} {\sqrt{\mbox{det}( X_0)}}   \exp [ \frac
{i m} {2 \hbar} (\mathbf{r}-\mathbf{r_0}) Y_0 X_0^{-1}
(\mathbf{r}-\mathbf{r_0}) \nonumber \\
& + \frac i \hbar (\mathbf{r}-\mathbf{r_0}) \cdot (\mathbf{p_0} -
2 \hbar \tilde{X_0^{-1}} \overrightarrow{\alpha} ) + \frac 1 2
\overrightarrow{\alpha} X_0^{-1} X_0^{*} \overrightarrow{\alpha} ]
\end{eqnarray}
The matrices $X_0,Y_0$ and the vectors $\mathbf{r_0},\mathbf{p_0}$
 correspond respectively to the position and momentum widths, to the average position and to the average momentum of
the wave function. We can then restrict our derivation without any
loss of generality to the propagation of
(\ref{eq:generatinggaussian}). Since the Hamiltonian
$H_{\mbox{G}}$ can be considered with a very good approximation to
be quadratic in position and momentum, the
wave-packet~(\ref{eq:generatinggaussian}) follows the ABCD law for
atom optics~\cite{BordeTheortool2001}. In order to alleviate the
notations, we shall omit to mention the index
$\overrightarrow{\alpha}$ in the subsequent computations and
denote the corresponding state
$|\Psi_{\mathbf{r_0},\mathbf{p_0},X_0,Y_0} \rangle$.

\subsection{Initial Expansion}

Before the trap shutdown, the condensate evolves under the
Hamiltonian $H=p^2/2m+H_{G}+H_{\mbox{Trap}}$. We remove the
gravitational term $H_{G}$ thanks to a unitary transform
$U_G(t,t_0)$:
\begin{equation}
|\Psi(t) \rangle = U_G(t,t_0) |\Phi(t) \rangle
\end{equation}
$U_G(t,t_0)$ represents the evolution of a quantum state under a
gravitational field. Performing this unitary transform is
equivalent to study the condensate in the non-inertial free
falling frame. The state $|\Phi(t) \rangle$ then evolves under the
Hamiltonian $H_{Free Fall}=p^2/2m+H_{\mbox{Trap}}$. The condensate
is taken to be initially in the strong coupling regime, so that
the corresponding wave function $\Phi(\mathbf{r},t)$ follows the
scaling laws established by Castin and Dum~\cite{CastinDum96} for
the Thomas-Fermi expansion. The initial wave function
$\Phi(\mathbf{r},t_0)$ corresponds to the Thomas-Fermi profile. We
represent concisely its evolution by the unitary transform
$U_{TFE} (t,t_0)$:
\begin{equation}
|\Phi(t) \rangle = U_{TFE} (t,t_0) |\Phi(t_0) \rangle
\end{equation}
It is indeed not useful at this point to explicit this transform,
whose expression is given in Appendix~\ref{sec:appendix_Thomas
Fermi expansion}. We transform back to the laboratory frame at the
time $t_1$ when we start to shine the first ``mirror pulse'':
\begin{equation}
|\Psi(t_1) \rangle = U_G (t_1,t_0) U_{TFE} (t_1,t_0)
 |\Phi(t_0) \rangle
\end{equation}
The resulting quantum state $|\Psi(t_1) \rangle$ will be taken as
a starting point for the subsequent oscillation of the condensate
in the cavity. Its evolution is conveniently obtained by
decomposition on a suitable basis of Hermite-Gauss modes $H_{lmn}
(\mathbf{r})$, as explained in the preceding paragraph. The
initial free fall simply determines the initial coefficients of
the projection:
\begin{equation}
\label{eq:initial Thomas Fermi wavefunction}
\hat{\Psi}(\mathbf{r},t_1)= \sum_{l,m,n} c_{lmn}(t_1) H_{lmn}
(\mathbf{r})
\end{equation}

\subsection{Propagation of a Gaussian wave-packet in the diluted
regime}

%It is a priori not obvious how to decouple, in the evolution of
%the wave-packet, the effects of gravity and electromagnetic
%fields. During the Raman pulse, the acceleration of the atoms changes the
 The evolution of free falling atoms in a Raman pulse is non-trivial since the gravitational acceleration makes the detuning
 (\ref{eq:OffBraggparam1}) time-dependent:
\begin{eqnarray}
\Delta(t)= \omega_1- \omega_2 - \omega_{ba} - 2 (\mathbf{p}
 - m g \:t\:\mathbf{u}_z) \cdot \mathbf{k}
 / m \nonumber \\ - 2 \hbar\mathbf{k}^{2} /  m -
\delta_{AC}
\end{eqnarray}
The behavior of a two-level atom falling into a laser wave has
been solved exactly~\cite{BordeLam95}. Gravitation alters
significantly the two-level atom state trajectory on the Bloch
sphere when the pulse duration $\tau$ becomes of the order of:
\begin{equation}
\tau_{g}=\frac {1} {\sqrt{||\mathbf{k}|| ||\mathbf{g}||}} \simeq
10^{-4} s
\end{equation}
Indeed, for this duration the off-Braggness
parameter~(\ref{eq:OffBraggparam1}) changes significantly during
the pulse. As seen in the previous section, in order to probe
effectively the resonance condition~(\ref{eq:resonanceT}), the
Raman $\pi$-pulses need to be velocity-selective. This leads us to
consider pulse durations on the order of the millisecond,
typically longer than $\tau_{g}$. It is then necessary to
compensate the time-dependent term induced by the
acceleration of gravity in the detuning  by an opposite frequency ramp chirping the pulse.\\

The simultaneous effects of gravito-inertial and electromagnetic
fields can be decoupled thanks to an effective propagation scheme
developped by Antoine and
Bord\'{e}~\cite{BordeTheortool2001,BordeMetrologia2002}. It
accounts for the electromagnetic interaction through an
instantaneous diffusion matrix $\hat{S}$ and for gravito-inertial
effects through a unitary transform $U_1(T,0)$:
\begin{equation}
|\Psi(T)\rangle = U_1 (T,0) \: \hat{S} \: |\Psi(0)\rangle
\end{equation}
Following this propagation method, it is sufficient to apply an
effective instantaneous diffusion matrix for each ``mirror pulse''
and evolve the state between the pulse centers as if there was no
electromagnetic field.
%Indeed we will ignore the dispersion effects in the action of the
%diffusion matrix $\hat{S}$ on the gaussian wave-function.

\subsection{Action of the effective instantaneous interaction matrix
$\hat{S}$}

We study in this paragraph the interaction of the condensate with
a quasi-plane electromagnetic wave, for which the instantaneous
diffusion matrix $\hat{S}$ is known. This matrix is
operator-valued, but momentum operators can be taken as
complex-numbers since the considered wave function is a narrow
momentum wave-packet centered around a nearly resonant momentum
$\mathbf{p_0}$ (i.e. such that $y(p_0) \ll 1$). In other
circumstances, this interaction can give rise to fine structuring
effects such as the splitting of the initial wave into several
packets following different trajectories (Borrmann
effect)~\cite{BordeBerman}. Following the approach of the
paragraph~\ref{subsec:condensate description}, we consider a
Gaussian matter-wave:
\begin{equation}
|\Psi_0 \rangle = |a,\Psi_{\mathbf{r_0},\mathbf{p_0},X_0,Y_0}
\rangle
\end{equation}
We study the interaction of this atomic wave with a ``mirror
pulse'' involving two linearly polarized running laser waves:
\begin{eqnarray}
\label{eq:champpulse1} & \textbf{E}  =   \textbf{E}(x,y) \cos
\left(
k z - \omega_1 t + \Phi_1 \right) \nonumber \\
& +  \textbf{E}(x,y) \cos \left( k
z + \omega_2 t + \Phi_2 \right) \nonumber\\
\end{eqnarray}
With respect to the population transfer, electromagnetic fields
may be treated as plane waves in the vicinity of the beam waist.
 The effective diffusion matrix $\hat{S}(\mathbf{k}, \mathbf{p_0})$
associated to a Raman pulse effective wave vector $\mathbf{k}$ and
applied to a wave-packet of central momentum $\mathbf{p_0}$ then
yields~\cite{BordeBerman}:
\begin{eqnarray}
\hat{S}(\mathbf{k}, \mathbf{p_0}) =  \left(
\begin{array}{cc}
S_{bb} & S_{ba}  \\
S_{ab} & S_{aa}  \\
\end{array} \right)
%\begin{array}{c}
%|a,\Psi_{\mathbf{r_0},\mathbf{p_0},X_0,Y_0} \rangle \\
%|b,\Psi_{\mathbf{r_0},\mathbf{p_0+\hbar \mathbf{k}},X_0,Y_0}
%\rangle
%\end{array}
\end{eqnarray}

\begin{eqnarray}
\label{eq:matrix S}
 S_{aa} = S_{bb}^{*}=
\exp{\left[-  i
 (\Omega_{AC}^{e}(\mathbf{r})+\Omega_{AC}^{g}(\mathbf{r})) \frac \tau 2 \right]}
\exp{(i \delta_{12} \tau)
 } \nonumber \\
 \times \left[ \cos \left(\Omega_0 \tau \sqrt{1+y^2} \right) + i \frac
{2y} {\sqrt{1+4y^2}} \sin \left( \Omega_0 \tau \sqrt{1+y^2}
\right)
\right] \nonumber \\
 S_{ab} = S_{ba}= i
\exp{\left[-  i
 (\Omega_{AC}^{e}(\mathbf{r})+\Omega_{AC}^{g}(\mathbf{r})) \frac \tau 2 \right]}
\exp{(-i \delta_{12} \tau) } \nonumber \\
\times \sin \left( \Omega_0 \tau \sqrt{1+y^2} \right) /
 \sqrt{1+y^2}  \nonumber \\
y(\mathbf{p_0},\mathbf{k})= \frac
{-\delta_{12}(\mathbf{p_0},\mathbf{k})} {2\Omega_0} = - \frac {1}
{2\Omega_0} [ \omega_1- \omega_2 - \omega_{ab} - \mathbf{k}
\cdot \mathbf{p_0} / m  \nonumber \\
 - \hbar \mathbf{k}^{2} / 2 m - (\Omega_{AC}^{e}(\mathbf{r})-\Omega_{AC}^{g}(\mathbf{r}))
 ]\nonumber \\
\end{eqnarray}
$\omega_1,\omega_2$ are respectively the pulsations of the lasers
propagating upward and downward, $\tau$ the duration of the pulse,
$\Omega_{AC}^{e}(\mathbf{r})$ and $\Omega_{AC}^{g}(\mathbf{r})$
are the AC Stark shifts of the associated levels. It is worth
commenting the position dependence of those terms, which intervene
in two different places in the $S$ matrix. In the off-Braggness
parameter $y$, the term $\delta_{AC}(\mathbf{r})$ induces an
intensity modulation, while in the complex exponential, the term
$\Omega_0 (\mathbf{r})=
\Omega_{AC}^{e}(\mathbf{r})+\Omega_{AC}^{g}(\mathbf{r})$ changes
the atomic wave-front.

After each ``mirror pulse'', the part of the condensate which does
not receive the double momentum transfer will fall out of the trap
if $y(p_0+ m g T) \gg 1$. We thus project out those states and
focus on the non diagonal terms of the diffusion matrix:
\begin{equation}
|\Psi (2 \tau) \rangle = \langle a |
\hat{S}(-\mathbf{k},\mathbf{p_0}+\hbar \mathbf{k}) | b \rangle
\langle b | \hat{S}(\mathbf{k},\mathbf{p_0}) |\Psi(0) \rangle
\end{equation}
The state after the mirror pulse is thus:
\begin{eqnarray}
|\Psi (2 \tau) \rangle = \rho(\mathbf{r})
|a,\Psi_{\mathbf{r_0},\mathbf{p_0}+2 \hbar \mathbf{k},X_0,Y_0}
\rangle \nonumber \\
\rho(\mathbf{r}) = \frac {e^{-i
2(\delta_{12}-\Omega_{AC}^{0}(\mathbf{r})) \tau} \sin^{2} \left(
\Omega_0 \tau \sqrt{1+y(p_0)^2} \right)} {(\Omega_0 \tau
\sqrt{1+y(p_0)^2})^{2}}
\end{eqnarray}
The amplitude factor $\rho(\mathbf{r})$ reflects both the loss of
non-reflected atoms and the change in the atomic beam wave-front.
Expanding the generating wave
function~(\ref{eq:generatinggaussian}) into powers of $\alpha$,
one shows that the effect of the interaction matrix $S$ is the
same on each mode of the expansion~(\ref{eq:initial Thomas Fermi
wavefunction}).

\subsection{Gravito-inertial effects}

The unitary transform $U_1(T,0)$  represents the gravito-inertial
effects. We refer the interested reader
to~\cite{BordeMetrologia2002} for a thorough derivation of this
operator. We remind here the main result necessary for our
computation. This operator maps a state defined by a Gaussian of
parameters $X_1,Y_1,r_1,p_1$ in position representation onto an
other Gaussian state in position representation whose parameters
$X_2,Y_2,r_2,p_2$ depend linearly on the former according to:
\begin{equation}
\label{eq:gravity_ABCD law}
\left( \begin{array}{c} \mathbf{X_2} \\
\mathbf{Y_2}
\end{array}
\right)
 = \left(
\begin{array}{cc}
\cosh[\sqrt{\gamma}T] & \gamma^{-1/2} \sinh[\sqrt{\gamma}T]  \\
\gamma^{1/2} \sinh[\sqrt{\gamma}T]  & \cosh[\sqrt{\gamma}T]  \\
\end{array} \right)
\left( \begin{array}{c} \mathbf{X_1}\\
\mathbf{Y_1}
\end{array}
\right)
\end{equation}
The coefficient $\gamma$ reflects the interaction effects through
an effective potential quadratic in position and assumed to be
constant in time. There is the same matrix relation between the
initial and final position and momentum centers
$\mathbf{r_1},\mathbf{p_1},\mathbf{r_2},\mathbf{p_2}$ of the
wave-packets, with an additional function $\xi$ which reflects the
constant part of the gravity field. The transform $U_1(T,0)$ also
introduces an additional phase factor given by the classical
action:
\begin{equation}
U_1(T,0) |a,\Psi_{\mathbf{r_1},\mathbf{p_1},X_1,Y_1} \rangle =
e^{iS_{Cl}(T,0)} |a,\Psi_{\mathbf{r_2},\mathbf{p_2},X_2,Y_2}
\rangle
\end{equation}
Since this phase factor does not play any role in the following
computations, we do not give its expression here, but it can be
found in reference~\cite{BordeTheortool2001}. Expanding a generic
Gaussian such as (\ref{eq:generatinggaussian}) shows that the
propagation of any Hermite mode of the expansion~(\ref{eq:initial
Thomas Fermi wavefunction}) is identical in the gravity field: the
Gaussian parameters ${X,Y}$ involved in each mode are transformed
identically.

\subsection{Conclusion: cycle evolution of the matter wave}

In our approach, the effect of the interactions has been neglected
after the first bounce and the propagation of the diluted atomic
sample in the cavity is essentially mode-independent. Nonetheless,
in experiments where atomic samples of higher density are bouncing
on electromagnetic
mirrors~\cite{Bouyer93,Lewenstein99,Boshier02,Hinds99},
interactions do change the shape of the cloud during the
propagation. As we shall see in the next section, interactions
impact the transverse velocity distribution in a way that can lead
to a reduced stability of the cavity. In the following, we will
review possible focusing techniques to solve this problem.

%In principle, we could choose any initial parameters
%${p_0,X_0,Y_0}$ to define the projection basis of the modes
%$G_{lmn,X_0,Y_0,r_0,p_0} (\mathbf{r})$. However, in order to
%minimize the number of significantly excited modes, we shall adopt
%initial parameters ${p_0,X_0,Y_0}$ such that the Thomas-Fermi
%wave-function and the fundamental gaussian mode have identical
%momentum average  and dispersion.

\section{Matter-wave focusing}
\label{sec:refocusing}

We investigate in this section two possible curved mirrors. We
first review a focusing technique based on the phase imprinting
through a position-dependent Stark shift~\cite{Whyte04}.
Afterwards, we introduce an original focusing mechanism based on a
laser wave-front curvature transfer.

\subsection{Matter-wave focusing with phase imprinting}

This method has the advantage of leading to tractable equations.
It relies on a position dependent Stark shift provided by
quasi-plane waves with a smooth intensity profile:
\begin{equation}
\textbf{E}(x,y)= E_0 \left(1- \frac {x^2+y^2} {w^2} \right)
\mathbf{u}
\end{equation}
Unfortunately, this Stark shift implies a position-dependence due
in the population transfer. This results in a loss of atoms which
makes this focusing method hardly compatible with the extreme
cavity stability required by this experiment. Nonetheless, it is
interesting to demonstrate the effect on the wave curvature
induced by this position dependent light shift. In this
perspective, we neglect the position dependence in the population
transfer but not in the phase of the diffracted matter wave.
Indeed this shift $\Omega_{AC}(\mathbf{r})=  \Omega_{AC}^{0}
\left( 1- 2 \frac {x^2+y^2} {w^2} \right)$ imprints a quadratic
phase to the matter wave:
\begin{equation}
|\Psi (2 \tau) \rangle =  \rho_1 \exp[-i  \frac {4 \Omega_{AC}^{0}
\tau} {w^2} (x^2+y^2)] |a,\Psi_{\mathbf{r_0},\mathbf{p_0}+2 \hbar
\mathbf{k},X_0,Y_0} \rangle
\end{equation}
The position-independent phase shift is hidden in the coefficient
$\rho_1$. Using expression~(\ref{eq:generatinggaussian}) for the
wave function $\Psi_{\mathbf{r_0},\mathbf{p_0}+2 \hbar
\mathbf{k},X_0,Y_0}$, one can recast the last equation into:
\begin{equation}
|\Psi (2 \tau) \rangle = \rho_1
|a,\Psi_{\mathbf{r_0},\mathbf{p_0}+2 \hbar \mathbf{k},X_1,Y_1}
\rangle
\end{equation}
with:
\begin{equation}
\label{eq:transfoX1Y1}
\left( \begin{array}{c} \mathbf{X_1} \\
\mathbf{Y_1}
\end{array}
\right)
 = \left(
\begin{array}{cc}
1 & 0 \\
- \frac {8 \hbar \Omega_{AC}^{0} \tau} {m w^2}\: P_{\bot} & 1 \\
\end{array} \right)
\left( \begin{array}{c} \mathbf{X_0} \\
\mathbf{Y_0}
\end{array}
\right)
\end{equation}
with $P_{\bot}$ projection matrix on the transverse directions:
\begin{equation}
P_{\bot}= \left( \begin{array}{ccc} 1 & 0 & 0\\
                                    0 & 1 & 0\\
                                    0 & 0 & 0 \end{array}
\right) \nonumber
\end{equation}
The AC shift factor thus changes the Gaussian parameters of the
matter wave just like a thin lens of focal $f$ in classical
optics, where the transform law yields:
\begin{equation}
\label{eq:transfoX1Y1rayoptics}
\left( \begin{array}{c} \mathbf{X_1} \\
\mathbf{Y_1}
\end{array}
\right)
 = \left(
\begin{array}{cc}
1 & 0 \\
- \frac {1} {f} & 1 \\
\end{array} \right)
\left( \begin{array}{c} \mathbf{X_0} \\
\mathbf{Y_0}
\end{array}
\right)
\end{equation}
One could thus define the focal length of an atom optic device as
the $f$ parameter entering the ABCD transform
(\ref{eq:transfoX1Y1}). Precisely, a phase factor $\exp{\left[-i
\alpha (x^2+y^2+z^2)\right]}$ on a Gaussian atomic wave changes
the Gaussian parameters of the matter wave according to the ABCD
law of a thin lens of focal:
\begin{equation}
\label{eq:equivalent_focal} f = \frac {m} {2 \hbar \alpha}
\end{equation}
The AC dependent Stark shift thus plays for the atomic beam the
role of a thin lens of focal $f= m w^2 / 8 \hbar \Omega_{AC}^{0}
\tau $. Let us point our that this focusing occurs in the time
domain so that the ``focal length'' is indeed a duration.

\subsection{Matter-wave focusing with spherical light waves}

As mentioned in the last paragraph, the impossibility to maintain
a perfect population transfer on the whole wave-front while
focusing with a light shift effect makes this technique inadequate
for the proposed gravimeter. We investigate here a different
method which does not have this major drawback. Instead of shaping
the atomic wave-front thanks to an indirect light-shift effect, a
better way to proceed is indeed to have the
 matter wave interact with a light wave of suitable
wave-front, like on Figure~\ref{fig:interaction gaussian matter
light}.
\begin{figure}[htbp]
\begin{center}
\includegraphics[width=8.4cm]{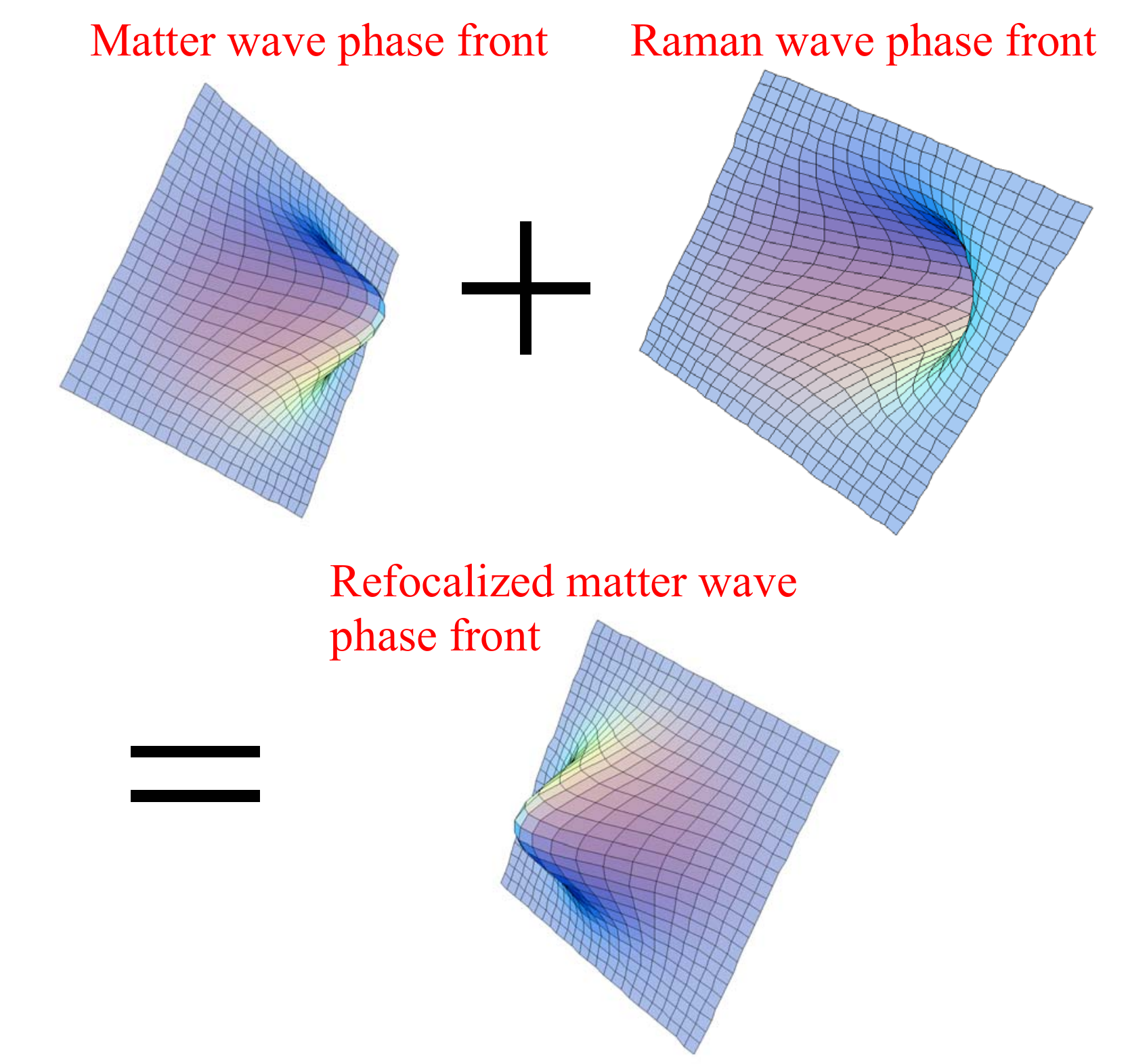}
\end{center}
\caption{Interaction between a spherical matter wave and a laser.
The laser spherical wave-front refocuses the matter
wave.}\label{fig:interaction gaussian matter light}
\end{figure}
Following up this intuitive picture, we propose an alternative
matter-wave focusing scheme, fully original to our knowledge,
based on the matter-wave interaction with electromagnetic fields
of spherical wave-front. To show that the focusing is effective,
we compute the transition amplitude of a matter wave interacting
with Gaussian Raman waves to first order in the electromagnetic
field. A similar computation has been previously performed by
Bord\'{e} in the context of atomic
beamsplitters~\cite{BordeGRG2004}.

 Let us consider for the ``mirror pulse'' two counterpropagating matched Gaussian beams:
\begin{eqnarray}
& E_1(x,y,z,t)= \frac {1} {2} U_0^{+}(\mathbf{r}-\mathbf{r_w})
e^{i k_0 (z-z_w)} E(t) e^{i \omega_1 t+i \varphi_1}+ c.c
\nonumber \\
& E_2(x,y,z,t)= \frac {1} {2} U_0^{-}(\mathbf{r}-\mathbf{r_w})
e^{-i k_0 (z-z_w)} E(t)e^{i \omega_2
t +i \varphi_2}+ c.c \nonumber \\
& \mbox{with} \quad U_0^{\pm}(\mathbf{r})=\frac {1} {1 \mp 2iz/b}
\exp [- \frac {1} {1 \mp 2iz/b}
\frac {x^2+y^2} {w_0^2} ]  \nonumber \\
\end{eqnarray}
where $k_0=(k_2-k_1)/2$. The detuning $\omega_2-\omega_1$ is
adjusted so that the relevant Raman process be the absorption of a
photon from mode $E_2$ followed by the emission of a photon into
mode $E_1$. We have used the confocal parameter of the light beam
$b=k_0 w_0^2$ as well as the complex Lorentzian function
$L^{+}$~\cite{Borde76}:
\begin{equation}
L^{+}(z)= \frac {1} {1-2 i z / ( k_0 w_0^2 )}=\frac {1} {\sqrt{ 1+
\frac {z^2} {4 b^2}
  }} \exp [ i \arctan \left( \frac
{z} {2 b} \right) ]
\end{equation}
The term $\arctan \left( \frac {z} {2 b} \right)$ is known as the
Gouy phase. The matching of the two laser beams is reflected in
the relation between their transverse structures
$U^{-}(\mathbf{r})=U^{+*}(\mathbf{r})$: at each point their
curvature is identical. The Raman diffusion associated with these
fields yields an effective interaction Hamiltonian whose matrix
elements is:
\begin{eqnarray}
\label{eq:potential two gaussian waves 1}
V_{ba}(\mathbf{r},t)=-\hbar \Omega &
U_0^{+}(\mathbf{r}-\mathbf{r_w})U_0^{-*}(\mathbf{r}-\mathbf{r_w})
e^{i 2 k_0 (z-z_w)} F(t)\nonumber \\ & \times   e^{-i(\omega_{21}+r(t-t_r))
t + i\varphi_0}+c.c.
\end{eqnarray}
The term $r(t-t_r)$ accounts for the frequency ramp starting at
time $t_r$, $\omega_{21}=\omega_2-\omega_1$ is the Raman detuning
and $F(t)=|E(t)|^2$ the time envelope of the pulse. The
computation of the transition amplitude is somewhat involved and
deferred to Appendix~\ref{sec:appendix_computation of the
first-order transition}. We obtain:
\begin{eqnarray}
 b^{(1)}(\mathbf{r},t)=  i \Omega \sqrt{2
\pi} e^{i 2 k_0 [z-z_{C0}(t)- 2 \hbar k_0 (t-t_0) /m]} e^{i
\varphi'_0} \nonumber \\ \times L^{+}(z-z_{C0}(t))
U_0^{+2}\left(\mathbf{r}-\mathbf{r_{C0}}(t)-\frac {2 \hbar
\mathbf{k_0}} {m} (t-t_0)  \right) \nonumber
\\ \times \langle b, \mathbf{r} |U_{0}(t,t_0) \int \frac
{d^{3}\mathbf{p}} {(2 \pi \hbar)^{3/2}}
  \left[ \tilde{F}(\omega_{B}(p_z,\mathbf{k_0}))
\langle
a, \mathbf{p}|\Psi(t_0)  \rangle \right]   |b, \mathbf{p} \rangle  \nonumber  \\
\end{eqnarray}
$U_{0}(t,t_0)$ is the evolution operator in the gravitational
field, $\mathbf{r}_{C0}(t)= \mathbf{r_w}+\frac {\mathbf{p_0}} {m}
(t-t_0)+ \frac 1 2 \mathbf{g} (t-t_0)^2$ and
$\omega_{B}(p_z,\mathbf{k_0})$ a frequency which reflects the
Bragg resonance condition:
\begin{equation}
\omega_{B}(p_z,\mathbf{k_0})=-[\omega_{ba} + \frac {\mathbf{k}
\cdot \mathbf{p}} {m} + \frac {\hbar \mathbf{k}^2} {2 m} -
\omega_{21} +\mathbf{k} \cdot \mathbf{g} \: (t_r-t_0)]
\end{equation}
$U_0^{+2}(\mathbf{r})$ corresponds to a Gaussian mode of confocal
parameter $b=k_0 w_0^2$ and waist $w_0/\sqrt{2}$.  The first-order
term~(\ref{eq:amplitude5}) is the leading contribution to the
outgoing excited matter wave.  The filtering of the pulse acts as
expected through the Fourier envelope
$\tilde{F}\left(\omega_{B}(p_z,\mathbf{k_0})\right)$, significant
only for a small velocity class which can be tuned by the starting
time $t_r$ of the velocity ramp.  The operator $U_{0}(t,t_0)$
reflects the propagation in the gravitational field. The factor
$L^{+}( z -z_C(t))$ barely affects the longitudinal shape of the
atomic wave, without contributing to the average vertical
momentum. As discussed in
Section~\ref{sec:principle_gravity_determination}, the cavity
lifetime of the atomic sample does not depend on its longitudinal
profile. Therefore we do not need to worry about this
factor. \\

What matters is the transverse structure of this outgoing wave,
which corresponds to a focusing matter wave. As suggested in
Figure~\ref{fig:interaction gaussian matter light}, the curvature
of the Gaussian Raman wave has been transmitted from the laser
wave to the atomic wave through the term
$U_0^{+2}[\mathbf{r}-\mathbf{r_{C0}}(t)-\frac {2 \hbar
\mathbf{k_0}} {m} (t-t_0)]$. This term induces a quadratic
dependence of the phase on spatial coordinates:
 \begin{equation} U_0^{+2}(\mathbf{r})=\exp [- \frac {2k_0^2 w_0^2 + i 4 k_0 z}  {k_0^2 w_0^4 + 4
z^2} (x^2+y^2) ]
\end{equation}
Following the approach of the precedent paragraph, and the
relation~(\ref{eq:equivalent_focal}), this can be interpreted as a
thin lens effect. To express the corresponding focal, we introduce
the vector
$\tilde{\mathbf{r}}(t)=\mathbf{r}-\mathbf{r_{C0}}(t)-\frac {2
\hbar \mathbf{k_0}} {m} (t-t_0)$. For an atomic wave centered
around $z=z_a$ at the time $t$, the interaction with the light
field plays the role of a thin lens of focal $f(z_a,z_w,t)$ to
first order in the electromagnetic field:
\begin{equation}
\label{eq:focal spherical wavefronts}
 f(z_a,z_w,t) = \frac {m}  {\hbar}  \frac  {k_0^2 w_0^4 + 4
\tilde{z}(t)^2 } {8  k_0 \tilde{z}(t)}
\end{equation}
where the parameter $\tilde{z}(t)$ is:
\begin{equation}
\tilde{z}(t)=z_a- \left(\frac {p_{0z}+2 \hbar k_0} {m} (t-t_0)
 - \frac 1 2 g (t-t_0)^2\right) - z_w
\end{equation}
%The focusing effect disappears if happens to be stronger if the
%diffusion process occurs far from the beam waist centered around
%$z=z_1$.
This focal can thus be controlled by the relative position of the
laser waist and atomic wave-packet center. This first-order
computation shows how the laser beam curvature is transferred to
the matter beam wave-front and suggests that it is possible to
focus matter waves through Raman pulses with a spherical
wave-front in a controllable manner.
%\begin{equation}
% L^{+} \left( z-z_1-(\frac {p_z} {m}+\frac {\hbar b} {2 m w_0^2} ) t
%\right) =  \frac {\exp [ i \frac {1} {2 b} \arctan
%\left(z-z_1-(\frac {p_z} {m} + \frac {\hbar b} {2 m w_0^2}) t
%\right) ]}  {\sqrt{ 1+ \frac {1} {4 b^2} \left( z--z_1-(\frac
%{p_z} {m}+\frac {\hbar b} {2 m w_0^2} ) t \right)^2 }}
%\end{equation}
%which induces in the vicinity of $\tilde{z}(t)$ an additional
%momentum of $\frac {\hbar} {2b} \mathbf{u_z}$

\subsection{Transverse stability of the cavity}

The main threat to the cavity stability is indeed the interatomic
interactions which will push away the atoms of the sample from the
central axis of the cavity. It is indeed possible to approximate
these interactions with an effective lens. A detailed ABCD matrix
analysis of interaction effects is given in~\cite{ImpensABCDinteractions}.
Here we just consider that interactions induce an effective
quadratic potential represented by the diagonal matrix
$\gamma_{i}$.

In the precedent paragraphs, the focusing obtained is only
effective for the transverse directions. This is sufficient, since
the longitudinal spread of the atomic sample does not drive it out
of the laser beams. The transverse stability of the cavity is
entirely reflected in the temporal evolution of the Gaussian
parameters $X(t),Y(t)$. The final parameters are related to the
initial parameters by the ABCD matrix:
\begin{equation}
\label{eq:matrixstability1}\left(
\begin{array}{cc}
\cosh[\sqrt{\gamma_{i}}T] - \frac {P_{\bot}} f \sqrt{\gamma_{i}}
\sinh[\sqrt{\gamma_{i}}T] & {\gamma_i}^{-1/2} \sinh[\sqrt{\gamma_{i}}T]  \\
\sqrt{\gamma_{i}} \sinh[\sqrt{\gamma_{i}}T] - \frac {P_{\bot}} f
\cosh[\sqrt{\gamma_{i}}T] & \cosh[\sqrt{\gamma_{i}}T]  \\
\end{array} \right)
\end{equation}
As in~\cite{ImpensABCDinteractions}, one can use this input-output relation to
model the stability of the matter-wave cavity. For the diluted matter waves involved in this system, a slight curvature in the mirror is sufficient to stabilize transversally the atomic beam.
%Since the gravity gradient is diagonal, the matrix
%\ref{eq:matrixstability1}) is indeed a tensor product of three
%$2\times2$ matrices $M_x \bigotimes M_y \bigotimes M_z$
%corresponding to each direction $x,y$ and $z$. 
%Therefore the time
%evolution of the longitudinal and transverse Gaussian parameters
%is decoupled.
%Longitudinally, the matter-wave spreads linearly in the vertical
%direction, but this expansion is not a great concern since it does
%not drive the condensate out of the laser beams.

% The matter wave will be confined if each of the
%transversal $2\times2$ matrix has eigenvalues inferior to unity in
%modulus. This condition is equivalent to:
%\begin{equation}
%\label{eq:stability_condition} 0 \leq \cosh[\sqrt{\gamma_{i x}}T]-
%\frac {\sqrt{\gamma_{i x}} \sinh[\sqrt{\gamma_{i x}}T]} {2 f} \leq
%1
%\end{equation}
%and a similar condition holds for $\gamma_{i y}$. 

\section{Conclusion}

We would like to point out the physical insight provided by the
picture of a cavity in momentum space. Such a cavity is only
possible in atom optics since photons, whose velocity is fixed to
$c$, cannot be accelerated. In our system, the corresponding
momentum wave-packet oscillates between two well-defined values
(Fig. \ref{fig:altitude_momentum}), with a resonance observed for
the adequate time-spacing of the mirrors. We can push further the
analogy with an optical cavity. The force $m g$ is the speed of
the field in the momentum space. $4 \hbar k$ is the momentum
analog for the cavity length. The cycle period $T_0$ of the wave
propagating in momentum space is orders of magnitude longer than
the usual cycle time of a pulse in an optical cavity.  \\

Let us look again at the resonance condition~(\ref{eq:resonanceT})
 with this picture in mind. This relation corresponds in momentum space to:
\begin{equation}
\label{eq:resonance_momentum space cavity} T = L / c
\end{equation}
with the replacements $m g \rightarrow  c$ (propagation velocity
in momentum space) and $4 \hbar k \rightarrow L$ (distance in
momentum space). The usual resonance condition for an optical
cavity yields:
\begin{equation}
T = n L / 2 c \quad n \in N
\end{equation}
 The difference can be
explained by the fact that, unlike in an optical cavity where the
light goes back and forth between the mirrors, the ``way back'' in
momentum space from $-2 \hbar k$ to $2 \hbar k$ has to be provided
by the light pulse. This explains why the factor $2$ is absent in
the denominator of (\ref{eq:resonance_momentum space cavity}). The
integer $n$ is absent in the resonance
condition~(\ref{eq:resonance_momentum space cavity}) because we
considered only two-photon Raman pulses for the optical mirrors.
Indeed, for each period $T_n =n T_0$, the cavity becomes resonant
with mirror pulses based on $n$-photons processes. In order to
levitate, the atomic sample needs to receive from the light pulse
an adequate momentum transfer of $4 n \hbar \mathbf{k}$. This
momentum fixes the number of photons exchanged for each possible resonant period $T_n$.\\

In the proposed gravimeter, the momentum cavity is loaded with
short single atomic pulses well-localized in momentum space, since
the instantaneous velocity distribution is sharply peaked at any
time. This is the analog of a femtosecond pulse propagating in an
optical cavity. It would be interesting, however, to load the
cavity with a continuous flow of free falling atoms coming from a
continuous atom laser. At a fixed momentum, the contributions from
different times would sum-up and interfere, exactly like in a
Perot-Fabry interferometer. This system would then constitute to
our knowledge the first example of a momentum space cavity
continuously loaded with a matter-wave beam.\\

We have studied the levitation of an atomic sample by periodic
double Raman pulses.  In our system, the matter wave is trapped in
an immaterial cavity of periodic optical mirrors. For the adequate
time interspace between two pulses, the atomic sample is
stabilized and levitate for a long time. Thanks to the sensitivity
of the stabilization to this period, one obtains an accurate
determination of the gravitational acceleration. \\

In our approach, the system could be loaded with any atomic sample
describable by a macroscopic wave function. It is indeed not
necessary to impose an initial small velocity dispersion, since
the first mirror pulse will serve as a filter for a narrow
velocity-class, while the next pulses will serve as a probe. Many
aspects developed in this paper are still valid for a thermal
cloud. Nonetheless, Bose Einstein condensates are ideally suited
for this trap since matter-wave focusing is more efficient with a
single mode coherent source. In this paper we have considered only
$\pi$-pulses for the atom-light interactions. In fact, one could
consider other schemes, for example one could split each
$\pi$-pulse in two copropagating $\pi/2$-pulses separated by a
dark space resulting in a sequence of Ramsey-Bord\'{e}
interferometers~\cite{MultiArches}. Since the sensitivity
to gravitation is proportional to the area covered in space-time
by the interferometer, the optimal situation is obtained when the
copropagating $\pi/2$-pulses are separated by $T_0/2$. An
experimental realization of this proposal is planned with the
support of the Institut Francilien de
Recherche en Atomes Froids(IFRAF).\\

\section{Acknowledgements}

We are very grateful to A. Landragin for valuable discussions and
suggestions. This work is supported by CNRS, CNES, DGA, and ANR.

%If the flow begins at $t=0$, one then has:
%\begin{equation}
%\psi(p,t)=\sum_{n=0}^{|t / T_0|} \psi(p,t- n T_0) \exp^{i n \phi}
%\end{equation}
%where $\phi$ is the phase flow in momentum space during a cycle.
%Bon il faut juste calculer cette phase, qui est egale a celle que
%l'on obtient quand on propage le paquet pour un cycle dans
%l'espace des $x$.

\appendix

\section{Computation of the gravimeter sensitivity}
\label{sec:appendix_sensitivity computation}

After $n$ nearly-resonant cycles, the fraction of the cloud
preserved becomes:
\begin{equation}
R(T) = R(p_1)...R(p_n) \quad \mbox{with} \quad R(p)= \frac {\sin^4
\left( \frac {\pi} 2 \sqrt{1+y(p)^2} \right)} {(1+y(p)^2)^2}
\end{equation}
with expression~(\ref{eq:momentum_at_pulse_n}) for the momenta:
\begin{equation}
p_n  = - m g T/2+ (n-1) \times m g (T_0 - T )
\end{equation}
At resonance $T=T_0$, one would have $p_1=..=p_n=-2 \hbar k$ and
$y(p_1)=..=y(p_n)=0$. For a big number of cycles $n$, in the
vicinity of the resonance $T \simeq T_0$ one still has
$y(p_1),...,y(p_n) \ll 1$. The expression for the reflection
coefficients simplifies to:
\begin{equation}
R(p_i)=\frac {1} {(1+y^2(p_i))^2}+
O(y^4(p_i))=1-2y^2(p_i)+O(y^4(p_i))
\end{equation}
The fraction of atoms kept in the cloud can then be expressed as:
\begin{eqnarray}
\label{eq:logR1}\log \frac {1} {R(T)} = - \sum_{i=1}^{n} \log
\left( R(p_i)
\right) \simeq \sum_{i=1}^{n} \log \left(1+2 y^2(p_i) \right) \nonumber \\
\end{eqnarray}
We insert expression (\ref{eq:momentum_at_pulse_n}) for the
momentum in (\ref{eq:OffBraggparam2}) to derive an expression for
the off-Braggness parameter:
\begin{equation}
y(p_i) \simeq \frac {g (T-T_0) k}
 {\Omega_0} \times i
\end{equation}
The reflection coefficient becomes:
\begin{equation}
\label{eq:logR2} \log \frac {1} {R(T)} \simeq \sum_{i=1}^{n} \log
\left(1+ 2 i^2 \frac{g^2 (T-T_0)^2 k^2} {\Omega_0^2} \right)
\end{equation}
This sum may be approximated by an integral because $i \gg 1$:
\begin{equation}
\label{eq:logR3} \log \frac {1} {R(T)} \simeq \int_1^n dx \log
\left(1+ 2 x^2 \frac{g^2 (T-T_0)^2 k^2} {\Omega_0^2} \right)
\end{equation}
This integral can be performed analytically:
\begin{equation}
\label{eq:integral} \int dx \log (1+ a x^2) = - 2 x + 2 \frac
{\arctan (\sqrt{a} x)} {\sqrt{a}}+x \log (1+a x^2)
\end{equation}
We set $a= 2 g^2 (T-T_0)^2 k^2 / \Omega_0^2$, which verifies
$\sqrt{a} n \simeq y(p_n) \ll 1$. We can then use expression
$(\ref{eq:integral})$ in $(\ref{eq:logR3})$ and Taylor expand the
right hand side:
\begin{equation}
\label{eq:logR4} \log \frac {1} {R(T)} \simeq - 2 n + 2 [ n -
\frac {1} {3} a n^3]+n \: a n^2= \frac {1} {3} a n^3
\end{equation}
We have omitted the small term coming from the lower bound of the
integral. We finally obtain:
\begin{equation}
\label{eq:logR5} \log \frac {1} {R(T)}= \frac {2 g^2 k^2 (T-T_0)^2
n^3} {3 \Omega_0^2}
\end{equation}
Let the condensate perform $n$ bounces for a range of values of
$T$ close to the expected value $T_0$, and detect the number of
atoms in the cloud afterwards. If a relative variation $\epsilon$
can be tracked experimentally, we can bound the period $T_0$
between $T_1$ and $T_2$ such that $R(T_1)=R(T_2)=1-\epsilon$.
According to our previous computation:
\begin{equation}
|T_2-T_0|\simeq |T_1-T_0| \simeq \sqrt {\frac {3
\log(1/1-\epsilon)} {2}} \frac {\Omega_0} {g k } \frac {1}
{n^{3/2}}
\end{equation}
We infer the gravitational acceleration from the period $T_0$
thanks to relation (\ref{eq:resonanceT}), so that their relative
errors are related by:
\begin{equation}
|\frac {\Delta g} {g}| \leq |\frac {\Delta T} {T}| + |\frac
{\Delta v_r} {v_r}|
\end{equation}
with the recoil velocity $v_r=\hbar k/ m$. This gives the
following upper bound for the relative error on the gravitational
acceleration $g$:
\begin{equation}
\frac {|\Delta {g}|} {g}  \leq \sqrt {\frac {3} {8}} \frac {1}
{\hbar k^2} \left( \frac {\Omega_0 \sqrt{-\log(1-\epsilon)}}
{n^{3/2}} \right)+ |\frac {\Delta v_r} {v_r}|
\end{equation}

\section{Thomas-Fermi expansion} \label{sec:appendix_Thomas Fermi
expansion}

The evolution of a condensate initially in the strong coupling
regime yields~\cite{CastinDum96}:
\begin{equation}
 U_{TFE} (t,t_0) |\Phi(t_0) \rangle =  |\Phi(t) \rangle
\end{equation}
with:
\begin{eqnarray}
& \Phi(\mathbf{r},t)= \frac {e^{-i\beta(t)} e^{i m \sum_{j} r_j^2
\dot{\lambda}_j(t) / 2 \hbar \lambda_j(t)}} {\sqrt{\lambda_1(t)
\lambda_2(t) \lambda_3(t)}}
\tilde{\Phi} (\mathbf{r} / \lambda_j(t) ,t_0) \nonumber \\
& \mbox{and} \: \tilde{\Phi} (\mathbf{r},t_0) \simeq \left( \frac
\mu {N_0 g} \right)^{1/2} \left(1 - \omega_{\bot}^2 \frac
{(x^2+y^2)} {R^2} -\omega_{z}^2 \frac {z^2} {Z^2} \right)^{1/2}
\nonumber
\end{eqnarray}
 For a cigar-shaped condensate, the frequency ratio $\epsilon = \frac
{\omega_z} {\omega_{\bot}}$ is small and we may keep track of the
radial expansion only:
\begin{eqnarray}
\label{eq:Thomas_Fermi_scaling_parameters} \lambda_z(t)=1 \qquad
\lambda_{\bot}(t)=\sqrt{1+\omega_{\bot}^2t^2}\\
\beta(t)=\frac {\mu} {\hbar \omega_{\bot}} \arctan
\left(1+\omega_{\bot}^2t^2\right)
\end{eqnarray}

\section{Propagation of a wave function: the method of the
generating function} \label{sec:appendix_generating function
method}

Let us assume that we know the solution of a linear PDE for a
family of initial conditions indexed by $\overrightarrow{\alpha}$:
\begin{eqnarray}
& \partial_t f(\mathbf{r},t,\overrightarrow{\alpha})=
\L(f)(\mathbf{r},t,\overrightarrow{\alpha})\nonumber \\
& f(\mathbf{r},0,\overrightarrow{\alpha}) =  \frac {1}
{\sqrt{\mbox{det}( X_0)}}   \exp [ \frac {i m} {2 \hbar}
(\mathbf{r}-\mathbf{r_0}) Y_0 X_0^{-1}
(\mathbf{r}-\mathbf{r_0}) \nonumber \\
& + \frac i \hbar (\mathbf{r}-\mathbf{r_0}) \cdot (\mathbf{p_0} -
2 \hbar \tilde{X_0^{-1}} \overrightarrow{\alpha} ) + \frac 1 2
\overrightarrow{\alpha} X_0^{-1} X_0^{*} \overrightarrow{\alpha}
]=\Psi_{\overrightarrow{\alpha}} (\mathbf{r}) \nonumber \\
\end{eqnarray}
where $\L$ is a linear differential operator in the first two
variables of the function $f$. Hermite modes can be defined
through an analytic expansion of the exponential
$\Psi_{\overrightarrow{\alpha}} (\mathbf{r})$:
\begin{eqnarray}
\label{eq:Hermite_definition} &  \exp [ \mathbf{r} M_1 \mathbf{r}
+ \mathbf{r} M_2 \overrightarrow{\alpha} + \overrightarrow{\alpha}
M_3 \overrightarrow{\alpha}] \nonumber \\
& = e^{\mathbf{r} M_1 \mathbf{r}} \sum_{l,m,n} i^{l+m+n}
\alpha_1^{l} \alpha_2^{m} \alpha_3^{n} H_{lmn}( \tilde{M_2}
\mathbf{r}, - \frac 1 2 M_3 ) \nonumber \\
\end{eqnarray}
From the linearity of $\L$, the propagation of an Hermite-Gauss
mode $H_{lmn}$, i.e. the solution at future times of the partial
differential equation with initial condition:
\begin{eqnarray}
& \partial_t g_{lmn}(\mathbf{r},t)=
\L(g_{lmn})(\mathbf{r},t)\nonumber \\
& g_{lmn}(\mathbf{r},0) = H_{lmn}(  \mathbf{r} )
\end{eqnarray}
can be inferred from the coefficient of $\alpha_1^{l} \alpha_2^{m}
\alpha_3^{n}$ in the $\overrightarrow{\alpha}$ expansion of
$f(\mathbf{r},t,\overrightarrow{\alpha})$, after a change of
variables in the argument of the Hermite polynomial. The
propagation of any arbitrary wave function $\Psi(\mathbf{r},t)$
then follows by linearity from the computation of the initial
projection on the Hermite-Gauss basis:
\begin{eqnarray}
\Psi(\mathbf{r},t) = \sum_{l,m,n} c_{lmn} \: g_{lmn}(\mathbf{r},t)
\nonumber \\
\mbox{with} \quad  c_{lmn} = \int d^{3}\mathbf{r}
H_{lmn}^{*}(\mathbf{r}) \Psi(\mathbf{r},0)
\end{eqnarray}

\section{Computation of the first-order transition amplitude with spherical waves} \label{sec:appendix_computation of the first-order transition}

The state vector evolves under the Hamiltonian $H=H_0+H_E+V$,
where $H_0$ accounts for the internal atomic degrees of freedom,
$H_E=p^2 / 2 m + m g z$ for the external particle motion  and $V$
for the light-field.  In this appendix we compute the  transition
amplitude to first order in $V$. To perform this computation, we
consider the state vector $|\tilde{\Psi}(t) \rangle$ in the
interaction picture:
\begin{equation}
|\Psi(t) \rangle = U_{0}(t,t_1) |\tilde{\Psi}(t) \rangle
\end{equation}
$U_{0}(t,t_1)$ is the free evolution operator in the absence of
light field between times $t_1$ and $t$:
\begin{equation}
U_{0}(t,t_1)= \exp \left( - i H_E (t-t_1) / \hbar \right) \exp(- i
H_0 (t-t_1) / \hbar)
\end{equation}
 The light field is turned on at time $t_0$. The first-order term of the Dyson series associated with the potential $V$
is:
\begin{equation}
|\tilde{\Psi}^{(1)}(t) \rangle = \frac 1 {i \hbar} \int_{t_0}^{t}
dt' \tilde{V}(t') |\tilde{\Psi}^{(0)}(t_0) \rangle
\end{equation}
where $\tilde{V}(t')$ is the potential in the interaction picture:
\begin{eqnarray}
\tilde{V}(t)= U_{0}^{-1}(t,t_1) \left( V_{ba}(\mathbf{r_{op}},t)
\otimes |b \rangle \langle a | \right) U_{0}(t,t_1) \: + \: h.c.  \nonumber \\
=V_{ba}\left(\mathbf{R_{op}}(t,t_1),t\right) \otimes |b
\rangle \langle  a | e^{i \omega_{ba} (t-t_1)}  \: + \: h.c.  \nonumber \\
\end{eqnarray}
$\mathbf{R_{op}}(t,t_1)$ is the position operator in the
interaction picture, given by integration of the Heisenberg
equation of motion~\cite{BordeMetrologia2002}:
\begin{eqnarray}
\mathbf{R_{op}}(t,t_1)=U_{0}^{-1}(t,t_1) \mathbf{r_{op}}
U_{0}(t,t_1)
\nonumber \\
= A(t,t_1) \mathbf{r_{op}} + B(t,t_1) \mathbf{p_{op}} +
\mathbf{\xi}(t,t_1)
\end{eqnarray}
The parameter $t_1$,  associated with a choice of representation
for the interaction picture, can be chosen as $t_1=t_0$. We need
only consider the term $V_{ba}$ of the interaction potential, for
which we adopt the usual rotating wave approximation. The
first-order transition amplitude $b^{(1)}(\mathbf{r},t)$ is given
by the relation:
\begin{eqnarray}
\label{eq:amplitude0} b^{(1)}(\mathbf{r},t)= \frac 1 {i \hbar} &
\langle b, \mathbf{r} | U_{0}(t,t_0) \int_{t_0}^{t} dt'
V_{ba}\left(\mathbf{R_{op}}(t',t_0),t'\right)  \otimes |b \rangle
\langle a | \nonumber\\ & \times e^{i \omega_{ba} (t-t_0)}
U^{-1}_{0}(t_0,t_0) |\Psi(t_0) \rangle
\end{eqnarray}
In order to understand how the light wave structures the atomic
wave-packet, we introduce the matrix elements of $V$ between plane
atomic waves:
\begin{eqnarray}
\label{eq:amplitude1} b(\mathbf{r},t)=  \frac 1 {i \hbar} \langle
b, \mathbf{r} | U_{0}(t,t_0)
 \int_{t_0}^{t} dt' \frac {d\mathbf{p}
d\mathbf{p'}} {(2 \pi \hbar)^{3}} |b,\mathbf{p'} \rangle e^{i
\omega_{ba} (t'-t_0)} \nonumber
\\ \times \langle \mathbf{p'}| V_{ba}\left(\mathbf{R_{op}}(t',t_0),t'\right)
 |\mathbf{p} \rangle e^{i \omega_{ba} (t'-t_0)}\langle a, \mathbf{p}
|\Psi(t_0) \rangle \nonumber \\
\end{eqnarray}
 We introduce the Fourier transform of the interaction potential:
\begin{eqnarray}
\label{eq:Potential Fourier tranform} V_{ba}(\mathbf{r},t)=-\hbar
\Omega F(t)
e^{-i \frac r 2 (t-t_r)^2-i \omega_{21}(t-t_0)+i \varphi_0} \nonumber \\
\times \int \frac {d^3 \mathbf{k}} {(2 \pi)^{3/2}} W(\mathbf{k}) e^{i
\mathbf{k} \cdot (\mathbf{r}-\mathbf{r_w})} \nonumber \\
\end{eqnarray}
To compute the Fourier components of $V$, we first use the BCH
relation:
\begin{eqnarray}
& \: & e^{i \mathbf{k} \cdot (A(t',t_0) \mathbf{r_{op}} + B(t',t_0)
\mathbf{p_{op}} + \xi(t',t_0))}  
  =  e^{i \mathbf{k} \cdot A(t',t_0) \mathbf{r_{op}}} e^{i \mathbf{k} \cdot B(t',t_0)
\mathbf{p_{op}}} \nonumber \\
& \times & e^{\frac 1 2 [ \tilde{A}(t',t_0) \mathbf{k} \cdot
 \mathbf{r_{op}},
 \tilde{B}(t',t_0) \mathbf{k} \cdot
\mathbf{p_{op}} ]+ i \xi(t',t_0))} \nonumber \\
\end{eqnarray}
The last commutator is responsible for the recoil term, and can be
written:
\begin{equation}
\frac 1 2 [ \tilde{A}(t',t_0) \mathbf{k} \cdot \mathbf{r_{op}},
\tilde{B}(t',t_0) \mathbf{k} \cdot \mathbf{p_{op}}]= \frac {i
\hbar} 2 \tilde{\mathbf{k}} A(t',t_0) \tilde{B}(t',t_0) \mathbf{k}
\end{equation}
The matrix element of the interaction potential contains the
following term:
\begin{eqnarray}
\langle \mathbf{p'}|W(\mathbf{k}) e^{i \mathbf{k} \cdot (A(t',t_0)
\mathbf{r_{op}} + B(t',t_0) \mathbf{p_{op}} + \xi(t',t_0))} |
\mathbf{p} \rangle  \nonumber \\
= W(\mathbf{k}) e^{i \mathbf{k} \cdot B \mathbf{p}- i \frac {
\hbar} {2 m} \tilde{\mathbf{k}} A \tilde{B} \mathbf{k}+  i \xi}
\langle \mathbf{p'}| e^{i \tilde{A} \mathbf{k} \cdot
\mathbf{r_{op}}} | \mathbf{p} \rangle \nonumber
\\ = W(\mathbf{k}) e^{i \mathbf{k} \cdot B \mathbf{p}+ i \frac {
\hbar} {2 m} \tilde{\mathbf{k}} A \tilde{B} \mathbf{k}+ i \xi}
\delta(\mathbf{p'}-\mathbf{p}-\hbar \tilde{A} \mathbf{k})
\end{eqnarray}
we omitted the $(t',t_0)$ to alleviate the notations. The
transition amplitude~(\ref{eq:amplitude1}) becomes:
\begin{widetext}
\begin{eqnarray}
\label{eq:amplitude2}  b^{(1)}(\mathbf{r},t)=  i \Omega
  \int \frac {d^{3}\mathbf{p}
d^{3}\mathbf{k}} {(2 \pi \hbar)^{3/2} (2 \pi)^{3/2}}
\int_{t_0}^{t} dt'  \langle b, \mathbf{r} | U_{0}(t,t_0)
|b,\mathbf{p}+\hbar \tilde{A}(t',t_0) \mathbf{k} \rangle
W(\mathbf{k}) e^{i \mathbf{k} \cdot (B(t',t_0) \mathbf{p}+ i
\xi(t',t_0)-\mathbf{r_w})} \nonumber
\\ \times e^{i \frac {\hbar} {2 m} \tilde{\mathbf{k}} A(t',t_0) \tilde{B}(t',t_0) \mathbf{k}}
 e^{i \omega_{ba} (t'-t_0)} F(t') e^{-i \frac r 2 (t'-t_r)^2-i
\omega_{21}(t'-t_0)+i \varphi_0} \langle a, \mathbf{p}|\Psi(t_0)
\rangle
\end{eqnarray}
 The evolution of the $ABCD\xi$ parameters in a
constant gravitational field is simple:
\begin{equation}
A(t',t_0)=1 \quad B(t',t_0)= \frac {t'-t_0} m  \quad \xi(t',t_0)=
\frac 1 2 \mathbf{g} (t'-t_0)^2
\end{equation}
One can extend the computation to include the effect of
interactions by taking into account an effective lensing effect in
the $ABCD$ matrices. This approach will be developed elsewhere. We
introduce the Fourier transform $\tilde{F}(\omega)= \int \frac
{dt} {\sqrt{2 \pi}}  F(t) e^{-i\omega t}$ of the slowly varying
envelope $F(t')$. The phase in the integral~(\ref{eq:amplitude2})
is a second-order polynomial in $t'$:
\begin{eqnarray}
\varphi(t')=\frac 1 2 (\mathbf{k} \cdot \mathbf{g}-r){t'}^2
 +[ \omega_{ba} + \frac {\mathbf{k} \cdot \mathbf{p}} {m}  +  \frac {\hbar \mathbf{k}^2} {2 m}
- \omega_{21}+r\: t_r  -\mathbf{k} \cdot \mathbf{g} \:t_0 + \omega
] t' + \varphi'_0
\end{eqnarray}
where $\varphi'_0$ is a constant phase term. In order to maximize
the transition amplitude, the chirp rate $r$ should be adjusted to
cancel the quadratic variation of the phase, which yields as
anticipated in Section~\ref{sec:principle_gravity_determination}:
\begin{equation}
r = \mathbf{k} \cdot \mathbf{g}
\end{equation}
The time $t_r$, at which the frequency ramp begins, selects the
velocity class of the atoms which undergo the transition. Indeed,
the momentum $p$ of these atoms satisfies:
\begin{equation}
|\omega_{ba} + \frac {\mathbf{k} \cdot \mathbf{p}} {m}  + \frac
{\hbar \mathbf{k}^2} {2 m} - \omega_{21} +\mathbf{k} \cdot
\mathbf{g} \: (t_r-t_0) | \leq \Delta \omega
\end{equation}
where $\Delta \omega$ is the spectral width of the time envelope
$F(t')$. As a consistency check, we see that this condition
reproduces the resonance condition~(\ref{eq:resonanceBragg}) for
$t_r=t_0$.
 The next step in the computation
of the amplitude~(\ref{eq:amplitude2}) is to consider that the
matter wave is out of the interaction zone at the initial and
final times. This is legitimate, since we are in fact interested
in computing a scattering amplitude. This simplification allows us
to extend the bounds of the time integral to infinity, which
yields a Dirac distribution:
\begin{eqnarray}
\label{eq:amplitude3}  b^{(1)}(\mathbf{r},t)=  i \Omega \sqrt{2
\pi} e^{i \varphi'_0}
  \int \frac {d^{3}\mathbf{p}
d^{3}\mathbf{k}} {(2 \pi \hbar)^{3/2} (2 \pi)^{3/2}}
  \langle b, \mathbf{r} | U_{0}(t,t_0)
|b,\mathbf{p}+\hbar \mathbf{k} \rangle \nonumber
\\ \times
W(\mathbf{k}) e^{-i \mathbf{k} \cdot \mathbf{r_w}} \langle a,
\mathbf{p}|\Psi(t_0)  \rangle \int d\omega \tilde{F}(\omega)
\delta \left(\omega+\omega_{ba} + \frac {\mathbf{k} \cdot
\mathbf{p}} {m} + \frac {\hbar \mathbf{k}^2} {2 m} - \omega_{21}
+\mathbf{k} \cdot \mathbf{g} \: (t_r-t_0)\right)
\end{eqnarray}
The Bragg resonance condition selects the Fourier component of
adequate frequency in the temporal envelope. To alleviate the
notations, we note $\omega_{B}(\mathbf{p},\mathbf{k})$ the
frequency selected by the Bragg condition:
\begin{equation}
\omega_{B}(\mathbf{p},\mathbf{k})= \omega_{ba} + \frac {\mathbf{k}
\cdot \mathbf{p}} {m} + \frac {\hbar \mathbf{k}^2} {2 m} -
\omega_{21} +\mathbf{k} \cdot \mathbf{g} \: (t_r-t_0)
\end{equation}
To simplify the computation, it is useful to assume that the
spectrum $\tilde{F}(\omega)$ is broad enough to override
dispersion effects of the laser wave. In other words:
\begin{equation}
\omega_{B}(\mathbf{p},\mathbf{k}) \simeq
\omega_{B}(p_z,\mathbf{k_0})
\end{equation}
This is legitimate if the spectral width $\Delta \omega$ of the
pulse $F(t)$ verifies:
\begin{equation}
\Delta \omega \gg \Delta k_z \: \frac {p_{0}} m, \Delta k_{\bot}
\frac {\Delta {p_{\bot}}} m
\end{equation}
 Within these conditions, the Dirac integral leaves the amplitude:
\begin{eqnarray}
\label{eq:amplitude4}  b^{(1)}(\mathbf{r},t)=  i \Omega \sqrt{2
\pi} e^{i \varphi'_0}
  \int \frac {d^{3}\mathbf{p}
d^{3}\mathbf{k}} {(2 \pi \hbar)^{3/2} (2 \pi)^{3/2}}
  \langle b, \mathbf{r} | U_{0}(t,t_0)
|b,\mathbf{p}+\hbar \mathbf{k} \rangle W(\mathbf{k}) e^{-i
\mathbf{k} \cdot \mathbf{r_w}}
\tilde{F}(\omega_{B}(p_z,\mathbf{k_0})) \langle a,
\mathbf{p}|\Psi(t_0)  \rangle
\end{eqnarray}
 In order to see how  the curvature of the light beam is imprinted
onto the atomic beam, we need to perform the integration over
$W(\mathbf{k})$. To compute this spatial Fourier transform, we go
back to the expression of the potential (\ref{eq:potential two
gaussian waves 1}):
\begin{eqnarray}
V_{ba}(\mathbf{r},t)=-\hbar \Omega
U_0^{+2}(\mathbf{r}-\mathbf{r_w}) e^{i 2 k_0 (z-z_w)} F(t)
e^{-i(\omega_{21}+r(t-t_r)) t} +c.c.
\end{eqnarray}
where we have used the relation between the Gaussian modes
$U^{+}(\mathbf{r})=U^{-*}(\mathbf{r})$.
 The spatial function inside the potential
is defined by:
\begin{equation}
U_0^{+}(\mathbf{r})=\frac {1} {1 - 2iz/b} \exp \left[- \frac {1} {1 -
2iz/b} \frac {x^2+y^2} {w_0^2} \right]
\end{equation}
with the confocal parameter $ b = k_0 w_0^2$. It will be useful to
introduce its transverse Fourier transform:
\begin{equation}
U_0^{+}(\mathbf{r})= \frac {w_0^2} {4 \pi} \int dk_{x} dk_{y} \exp
\left[- \frac {(k_x^2+k_y^2) w_0^2} {4} (1 - 2iz/b) \right]
\end{equation}
It is convenient to introduce the Lorentzian
function~\cite{Borde76}:
\begin{equation}
L^{+}(z)=\frac {1} {1 - 2iz/b}
\end{equation}
The transverse Fourier transform of $W(\mathbf{r})$ can then be
expressed as:
\begin{eqnarray}
\label{eq:W(r)Fouriertransform}  W(\mathbf{r})={L^{+}}^2(z) \exp
\left[- 2 L^{+}(z) \frac {x^2+y^2} {w_0^2} \right] e^{i 2 k_0 z} 
= \int \frac {d^2 k_{\bot}} {(2 \pi)} \left(
 \frac {w_0^2} {4} L^{+}(z)   \exp \left[- \frac
{(k_x^2+k_y^2)} {4} \frac {w_0^2}
{2} (1 - 2iz/b) \right] e^{i 2 k_0 z} \right) e^{i k_{\bot} \cdot \mathbf{r_{\bot}}} \nonumber \\
\end{eqnarray}
%with $U_{2}^{+}(\mathbf{r})$ is the Gaussian mode associated with
%a smaller waist $w_0 / \sqrt{2}$ and identical confocal parameter.
From this last expression we infer:
\begin{equation}
\label{eq:additionalstepFouriertransform} \int \frac {d k_z} {(2
\pi)^{1/2}} W(\mathbf{k}) e^{i k z} =     L^{+}(z) \frac {w_0^2}
{4}
 \exp \left[- \frac {(k_x^2+k_y^2)} {4} \frac {w_0^2} {2} (1 -
2iz/b) \right] e^{+i 2 k_0 z}
\end{equation}
The propagation of a plane wave in a gravitational field yields:
\begin{equation}
\langle \mathbf{r} | U_{0}(t,t_0) |\mathbf{p} \rangle = \frac {1}
{\sqrt{2 \pi \hbar}} e^{\frac i \hbar \mathbf{p} \cdot \mathbf{r}}
e^{i S_{Cl}(t,t_0)}
\end{equation}
with $S_{Cl}(t,t_0)$ classical action between a trajectory of
initial momentum $\mathbf{p}$, final position $\mathbf{r}$ and
duration $t-t_0$. The corresponding expression can be recast as:
\begin{eqnarray}
\langle  \mathbf{r} | U_{0}(t,t_0) |\mathbf{p} \rangle = \frac {1}
{\sqrt{2 \pi \hbar}} e^{-i m g^2 (t-t_0)^3 / 6 \hbar} e^{- \frac
{i} {2 \hbar} \mathbf{p} \cdot \mathbf{g} (t-t_0)^2} 
e^{ \frac i \hbar (\mathbf{p}- m \mathbf{g} (t-t_0)) \cdot
\mathbf{r}}  e^{- i \mathbf{p}^2 (t-t_0)/2 m}
\end{eqnarray}
This gives an expression for the matrix element:
\begin{eqnarray}
\label{eq:propagation pesanteur} \langle b, \mathbf{r} |
U_{0}(t,t_0) |b,\mathbf{p}+\hbar
 \mathbf{k} \rangle = \frac {1} {\sqrt{2 \pi
\hbar}} e^{-i m g^2 (t-t_0)^3 / 6 \hbar} e^{- \frac {i} {2 \hbar}
\mathbf{p} \cdot \mathbf{g} (t-t_0)^2} \cdot e^{ \frac i \hbar
(\mathbf{p}- m \mathbf{g} (t-t_0)) \cdot \mathbf{r}}
 e^{ i \mathbf{k} \cdot (\mathbf{r}-\frac 1 2 \mathbf{g} (t-t_0)^2 -
\frac {\mathbf{p}} {m} (t-t_0) - \mathbf{r_w})} \nonumber \\
\times e^{-i\omega_b (t-t_0)}
 e^{- i \hbar \mathbf{k_z}^2/ 2 m (t-t_0)} e^{- i \mathbf{p}^2 (t-t_0)/2
m}  e^{- i \hbar \mathbf{k_{\bot}}^2 (t-t_0)/ 2 m}
\end{eqnarray}
 The momentum distribution
$W(\mathbf{k})$ peaked around the value $\mathbf{k}=2 k_0
\mathbf{u_z}$ has a width $\Delta \mathbf{k}\ll k_0$, and one can
verify on (\ref{eq:W(r)Fouriertransform}) that its longitudinal
width $\Delta k_z$
 is much narrower than the transverse ones $\Delta k_x, \Delta k_y$.
The correction to the longitudinal recoil $\hbar k_z^2 /2m$ when
$\mathbf{k}$ varies in the width of $W(\mathbf{k})$ is thus
typically much smaller than the transverse recoil. The term $\hbar
k_z^2 /2m$ in equation~(\ref{eq:propagation pesanteur}) will
therefore be approximated by $2\hbar k_0^2 /m$. By summing up the
Fourier modes, we will recover for the atomic wave the transverse
Fourier profile of the Gaussian laser wave
$(\ref{eq:W(r)Fouriertransform})$ up to a translation. We note
$\mathbf{r_c}(t)=\mathbf{r_w}+\frac {\mathbf{p}} {m} (t-t_0)+
\frac 1 2 \mathbf{g} (t-t_0)^2 $ the point associated with a
classical motion in the gravity field from $\mathbf{r_w}$. The
role played by the position associated with the classical movement
and the action phase pre-factor are indeed a consequence of the
$ABCD\xi$ theorem~\cite{BordeTheortool2001}. Gathering all the
terms of (\ref{eq:amplitude4}) dependent on the wave-vector
$\mathbf{k}$, and using
relation~(\ref{eq:additionalstepFouriertransform})we can perform
the integration on the wavevector $\mathbf{k}$ along:
\begin{eqnarray}
\label{eq:integration_on_k} & \: & \int \frac {d^{2}\mathbf{k}_{\bot}}
{(2 \pi)} \left[ \int \frac {dk_{z}} {(2 \pi)^{1/2}} W(\mathbf{k})
e^{i k_z (z-z_C)} \right] e^{i\mathbf{k_{\bot}} \cdot
(\mathbf{r_{\bot}}-\mathbf{r_{C\bot}})}
 e^{-i \frac {\hbar \mathbf{k_{\bot}}^2} {2 m} (t-t_0)} e^{-i \frac {2 \hbar k_0^2} {m} (t-t_0)}
\nonumber \\
& = & L^{+}(z-z_C(t)) e^{i 2 k_0 (z-z_C)} e^{-i \frac {2 \hbar k_0^2} {m}
(t-t_0)}  \frac {w_0^2} {4} \int \frac {d^{2}\mathbf{k}_{\bot}}
{(2 \pi)} \exp \left[- \frac {\mathbf{k_{\bot}}^2} {4} \frac {w_0^2}
{2} (1 - 2i(z-z_C)/b) \right] e^{-i \frac {\hbar \mathbf{k_{\bot}}^2} {2
m} (t-t_0)} e^{i\mathbf{k_{\bot}} \cdot
(\mathbf{r_{\bot}}-\mathbf{r_{C \bot}}(t))} \nonumber \\
 & = & L^{+}(z-z_C(t)) e^{i 2 k_0
[z-z_C-2 \hbar k_0 (t-t_0)/m]} \frac {w_0^2} {4} \int \frac
{d^{2}\mathbf{k}_{\bot}} {(2 \pi)} \exp \left[- \frac
{\mathbf{k}_{\bot}^2} {8} w_0^2 \left(1-\frac {2 i} {k_0 w_0^2}
(z-z_C(t)- \frac {2 \hbar k_0 (t-t_0)} {m}\right)\right]
e^{i\mathbf{k_{\bot}} \cdot (\mathbf{r_{\bot}} -\mathbf{r_{C
\bot}}(t) )}
\nonumber \\
& = & L^{+}(z-z_C(t)) e^{i 2 k_0
[z-z_C(t)-  \frac{2\hbar k_0} {m} (t-t_0)]}  U_0^{+2}\left(\mathbf{r}-\mathbf{r_{C}}(t)-\frac {2 \hbar \mathbf{k_0}} {m} (t-t_0)  \right) \nonumber \\
\end{eqnarray}
The momentum acquired during the Raman process is reflected in the
factor $e^{i 2 k_0 [z-z_C(t)-  \frac{2\hbar k_0} {m} (t-t_0)]}$.
The translation $\mathbf{r}-\mathbf{r_{C}}(t)-\frac {2 \hbar
\mathbf{k_0}} {m} (t-t_0) $ accounts for the classical motion in
the gravitational field and the momentum acquired during the Raman
process. Inserting this result in the
equation~(\ref{eq:amplitude4}):
\begin{eqnarray}
\label{eq:amplitude5}
 b^{(1)}(\mathbf{r},t) & = &  i \Omega \sqrt{2
\pi}  e^{i 2 k_0 [z-z_{C0}(t)- 2 \hbar k_0 (t-t_0) /m]} e^{i
\varphi'_0} \nonumber \\  & \times & \int \frac {d^{3}\mathbf{p}} {(2 \pi
\hbar)^{3/2}} e^{-i m g^2 (t-t_0)^3 / 6 \hbar} e^{- \frac {i} {2
\hbar} \mathbf{p} \cdot \mathbf{g} (t-t_0)^2} e^{- i \mathbf{p}^2
(t-t_0)/2 m}
 e^{ \frac i \hbar (\mathbf{p}- m \mathbf{g}
(t-t_0)) \cdot \mathbf{r}} e^{-i\omega_b t}
 \nonumber  \\
 & \times & L^{+}(z-z_C(t)) U_0^{+2}\left(\mathbf{r}-\mathbf{r_{C0}}(t)-\frac {2 \hbar
\mathbf{k_0}} {m} (t-t_0)  \right)
\tilde{F}(\omega_{B}(p_z,\mathbf{k_0})) \langle a, \mathbf{p}|\Psi(t_0)  \rangle  \nonumber  \\
\end{eqnarray}
If the atomic wave-packet is sufficiently narrow, the Gaussian
modes $L^{+}(z-z_C(t))
U_2^{+}\left(\mathbf{r}-\mathbf{r_{C}}(t)-\frac {2 \hbar
\mathbf{k_0}} {m} (t-t_0)  \right)$, which depend on the momentum
$\mathbf{p}$ through $r_{C}(t)$, are approximately constant on the
width of the distribution $\tilde{F}(\omega_{B}(p_z,\mathbf{k_0}))
\langle a, \mathbf{p}|\Psi(t_0)  \rangle $ centered on
$\mathbf{p_0}$. We can then pull those functions out of the
momentum integral. The phase factor in the momentum integral
correspond to the propagation of plane waves in a gravitational
field. One can thus interpret the momentum integral as the
propagation of the filtered wave-packet in the gravitational
field:
%\begin{equation}
%|\Psi'(t_0) \rangle = \lambda \int d\mathbf{p} |a, \mathbf{p}
%\rangle \tilde{F}(\omega_{B}(p_z,\mathbf{k_0})) \langle a,
%\mathbf{p}|\Psi(t_0)  \rangle
%\end{equation}
\begin{eqnarray}
 b^{(1)}(\mathbf{r},t)=  i \Omega \sqrt{2
\pi} e^{i 2 k_0 [z-z_{C0}(t)- 2 \hbar k_0 (t-t_0) /m]} e^{i
\varphi'_0} L^{+}(z-z_{C0}(t))
U_0^{+2}\left(\mathbf{r}-\mathbf{r_{C0}}(t)-\frac {2 \hbar
\mathbf{k_0}} {m} (t-t_0)  \right) \nonumber
\\ \times \langle b, \mathbf{r} |U_{0}(t,t_0) \int \frac
{d^{3}\mathbf{p}} {(2 \pi \hbar)^{3/2}}
  \left[ \tilde{F}(\omega_{B}(p_z,\mathbf{k_0}))
\langle a, \mathbf{p}|\Psi(t_0)  \rangle \right]   |b, \mathbf{p}
\rangle \nonumber
\end{eqnarray}
with $\mathbf{r}_{C0}(t)= \mathbf{r_w}+\frac {\mathbf{p_0}} {m}
(t-t_0)+ \frac 1 2 \mathbf{g} (t-t_0)^2$. To first order in the
field, the curvature of the Gaussian Raman wave is transferred in
a controlled way to the atomic wave through the terms
$L^{+}(z-z_{C0}(t))
U_0^{+2}\left(\mathbf{r}-\mathbf{r_{C0}}(t)-\frac {2 \hbar
\mathbf{k_0}} {m} (t-t_0)  \right)$.
\end{widetext}

\end{document}